\documentclass[12pt]{article}
\usepackage{fullpage}
\usepackage{epsf}
\usepackage{epsfig}
\usepackage{amsthm}
\usepackage{amsfonts}
\usepackage{color}
\usepackage{amsmath}
\usepackage{setspace}
\usepackage{natbib}
\usepackage{url}

\begin{document}

\newcommand{\bm}[1]{\mbox{\boldmath $#1$}}
\newcommand{\mb}[1]{\mathbf{#1}}
\newcommand{\bE}[0]{\mathbb{E}}
\newcommand{\bP}[0]{\mathbb{P}}
\newcommand{\ve}[0]{\varepsilon}
\newcommand{\Var}[0]{\mathbb{V}\mathrm{ar}}
\newcommand{\Corr}[0]{\mathbb{C}\mathrm{orr}}
\newcommand{\Cov}[0]{\mathbb{C}\mathrm{ov}}
\newcommand{\mN}[0]{\mathcal{N}}
\newcommand{\iidsim}[0]{\stackrel{\mathrm{iid}}{\sim}}
\newcommand{\NA}[0]{{\tt NA}}
\newcommand{\argmax}{\operatornamewithlimits{argmax}}

\title{\vspace{-1.7cm}Local Gaussian process approximation for large computer
  experiments}
\author{Robert B.~Gramacy\\
  Booth School of Business\\
  The University of Chicago\\
  {\tt rbgramacy@chicagobooth.edu} \and
  Daniel W.~Apley\\
  Dept of Industrial Engineering\\ and Management Sciences\\
  Northwestern University\\
  {\tt apley@northwestern.edu} 
}
 \date{}

\maketitle

\vspace{-0.5cm}
\begin{abstract}
  We provide a new approach to approximate emulation of large computer
  experiments.  By focusing expressly on desirable properties of the
  predictive equations, we derive a family of local sequential design
  schemes that dynamically define the support of a Gaussian process
  predictor based on a local subset of the data.  We further derive
  expressions for fast sequential updating of all needed quantities as
  the local designs are built-up iteratively.  Then we show how
  independent application of our local design strategy across the
  elements of a vast predictive grid facilitates a trivially parallel
  implementation.  The end result is a global predictor able to take
  advantage of modern multicore architectures, providing
  a nonstationary modeling feature as a bonus.  We
  demonstrate our method on two examples utilizing designs with
  thousands of data points, and
  compare to the method of compactly supported covariances.

  \vspace{-0.25cm}
\end{abstract}

\section{Introduction}
\label{sec:intro}

The Gaussian process (GP) is a popular choice for emulating computer
experiments \citep{sant:will:notz:2003}.  As priors for nonparametric
regression, they are unparalleled; they are rarely beaten in
out-of-sample predictive tests, have appropriate coverage, and have an
attractive ability to accomplish these feats while interpolating the
response if so desired.  That said, GPs do have disadvantages, for example: 
(1) computation (for inference and prediction) scales poorly as data sets
get large; and (2) the best results require an assumption of
stationarity in the data-generating mechanism, which may not be
appropriate.  The second issue is usually coupled with the first.
Emulators relaxing stationarity exist, but often
at further computational expense.  Examples include learning
nonlinear mappings to a space where stationarity
reigns \citep[e.g.][and references therein]{schmidt:ohagan:2003}, and
process convolution approaches that allow the kernels to vary smoothly
in parameterization as an unknown function of their spatial location
\citep[e.g.,][and therein]{paciorek:schervish:2006}.

Several works in recent literature have made inroads in reducing the
computational burden for stationary models, usually by making some
kind of approximation.  Examples include searching for a reduced set
of ``pseudo-inputs'' \citep{snelson:ghahr:2006}, building up
estimators iteratively \citep[e.g.,][]{haaland:qian:2012,gramacy:polson:2011},
fixed rank kriging
\citep{cressie:joh:2008}, compactly supported (sparse)
covariance (CSC) \citep{kaufman:etal:2012}.  Our contribution has
aspects in common with all of these approaches and several others by
association. 
It is reminiscent of {\em ad hoc} methods based on local kriging
neighborhoods \citep[e.g.,][pp.~131--134]{cressie:1993}. What we
propose is more modern, scalable, ideally suited to 
multicore computing, and tailored to computer
experiments. It draws in part on recent findings for approximate
likelihoods in spatial data \citep[e.g.,][]{stein:chi:welty:2004}, and
active learning techniques \citep[e.g.,][]{cohn:1996}.

We start by focusing on the prediction problem, locally, at an input
$x$.  In computer model emulation this is the primary problem of
interest.  We recognize, as many authors have before us, that data
points far from $x$ have vanishingly small influence on the predictive
distribution (assuming the usual choices of covariance).
Thus, even though a large covariance matrix may have been calculated
and inverted at great computational expense, not many of its
rows/columns contribute substantively to the linear predictor.  We
focus on finding those data points (relative to $x$), and
corresponding rows/cols, without considering the full matrices.  In
particular, we illustrate how a sensible objective criterion reduces
to a key component of an active learning heuristic that has found
recent application in the design of computer experiments literature
\citep[see, e.g.,][]{gra:lee:2009}.

Our localized sub-designs differ from a local nearest neighbor (NN) approach,
and yield more accurate predictions for a fixed local design size.  This
echoes a result by \cite{stein:chi:welty:2004} who observe that NNs may not be
ideal for learning correlation parameters.  We extend that to obtaining
predictions, although we recognize that a modern take on NN can represent an
attractive option, computationally, in many contexts. Our methods are
applicable with any covariance family that can be differentiated analytically.
However, we deliberately choose a very na\"ive one and nonetheless compare
favorably to others having more thoughtful choices
\citep[e.g.,][]{kaufman:etal:2012}.  The explanation is that we leverage a
globally nonstationary effect that compensates for an overly smooth/simplistic
local structure.

There are strong parallels between our suggested approach and CSC, although
there are two key distinctions.  The first is that our method is purely local,
whereas a substantial innovation in CSC is to take a local--global hybrid
approach.  CSC estimates a global mean structure via a polynomial basis in
order to ``soak up'' large scale variabilities in the spatial field.  
Second, we are not applying any kind of tapering of influence in learning that
local structure.  Tapering is a smooth operation
which would retain most of the features, e.g., stationary, of an un-tapered
analog.  However our approach is discrete and therefore could not retain
stationarity or smoothness.  Our main advantage is that
accuracy can be explicitly linked to a computational budget---we know exactly
how ``sparse'' our matrices will be before running the code, and we
can monitor convergence to the (intractable) full-data counterparts. Also,
since we focus on particular locations $x$, independent of others, our method
is highly parallelizable.  Local inference for many $x$'s
 can proceed without communication between nodes, meaning that
{\tt OpenMP} pragmas can yield nearly-linear speedups. 

The remainder of the paper is outlined as follows.  In Section
\ref{sec:gp} we review GP predictive modeling, inference and
prediction.  In Section \ref{sec:loc} we derive a 
criteria for greedily building local designs, expressions for the
efficient updating of all required quantities,
and argue that a simpler heuristic from the {\em
active learning} literature may prove to be even more efficient. 
Section \ref{sec:global} treats global prediction problem, for many
$x$, and discusses global inference for the correlation structure.  Finally, all
of the elements are brought together in Section \ref{sec:results} for
empirical illustrations and comparisons to CSC. We conclude with a discussion
in Section \ref{sec:discuss}.  An implementation of our methods
can be found in the {\tt laGP} \citep{laGP} for {\sf R}.

\section{Gaussian process predictive modeling}
\label{sec:gp}

A GP prior for functions $Y\!:\mathbb{R}^p \rightarrow \mathbb{R}$,
where any finite collection of outputs are jointly Gaussian,
is defined by its mean $\mu(x) = \bE\{Y(x)\}$ and
covariance $C(x,x') = \bE\{[Y(x) - \mu(x)][Y(x') - \mu(x')]^\top\}$.
A common simplifying assumption in the computer modeling literature is
to take the mean to be zero, and this is the convention we will
follow.  Typically, one separates out the variance $\tau^2$ in
$C(x,x')$ and works with correlations $K(x,x') = \tau^{-2} C(x,x')$
based on Euclidean distance and a small number of unknown parameters.
In this paper we focus on the isotropic Gaussian correlation
$K_{\theta,\eta}(x,x') = \exp\{ - ||x - x'||^2/\theta\} + \eta
\delta_{x,x'}$ where $\theta$ is called the {\em lengthscale}
parameter, and $\eta$ the {\em nugget}.  

In what follows we hold $\eta$ at a pre-determined small value in order to
obtain a near-interpolative surface which is appropriate when the computer
model is deterministic.  It is important to note that our main results do not
heavily depend on of the choice of correlation function so long as derivatives
(with respect to $\theta$) are available.  The isotropic, one-parameter
$(\theta)$ family simplifies the exposition. However the derivations of our
main results [Section \ref{sec:mspe}] assume a separable family. The {\tt
laGP} package facilitates estimating $\eta$ if desired.

Although technically a prior over functions, the regression
perspective recommends a likelihood interpretation.  Using data $D =
(X, Y)$, where $X$ is a $N\times p$ design matrix, the $N \times 1$
response vector $Y$ has a multivariate normal (MVN) likelihood with
mean zero and covariance $\Sigma(X) = \tau^2 K$, where $K$ is an $N
\times N$ matrix with $(ij)^{\mathrm{th}}$ entry $K(x_i,
x_j)$. Conditional on $K$, the MLE of $\tau^2$ is available
analytically.  Profile likelihoods may then be used to infer other
unknown parameters (e.g., $\theta$).

Empirical Bayes \citep{berger:deiliveira:sanso:2001} has gained in popularity
recently.  In this approach one specifies a reference prior,
$\pi(\tau^2)
\propto \tau^{-2}$. Having an MVN likelihood, and prior
algebraically equivalent to IG$(0,0)$, makes integrating over $\tau^2$
analytic.  We obtain
\begin{equation}
p(Y|K) = \frac{\Gamma[N/2]}{(2\pi)^{N/2}|K|^{1/2}} \times
\left(\frac{\psi}{2}\right)^{\!-\frac{N}{2}},
\label{eq:gpk}
\;\;\;\;\; \mbox{where} \;\;\;
\psi = Y^\top K^{-1} Y.
\end{equation}
Newton-like methods for estimating $\theta$ work
well when leveraging derivative information for the log likelihood
[see \ref{sec:mspe}], and when the likelihood surface is not multi-modal.

The marginalized predictive equations for GP regression, i.e., for
 $Y(x)$ given on data $D$ and covariance $K(\cdot,\cdot)$, are
Student-$t$ with degrees of freedom $N$,
\begin{align} 
  \mbox{mean} && \mu(x|D, K) &= k^\top(x)  K^{-1}Y,
\label{eq:predgp} \\ 
\mbox{and scale} && 
 \sigma^2(x|D, K) &=   
\frac{\psi [K(x, x) - k^\top(x)K^{-1} k(x)]}{N},
\label{eq:preds2}
\end{align}
where $k^\top(x)$ is the $N$-vector whose $i^{\mbox{\tiny th}}$
component is $K(x,x_i)$.  Using properties of the Student-$t$,
 the variance of $Y(x)$ is $V(x) \equiv \Var[Y(x)|D] =
\sigma^2(x|D,K)\times N/(N - 2)$.

\section{Localized approximate emulation}
\label{sec:loc}

If obtaining fast approximate prediction at $x$ is the goal, then a
first stab may be to use a small sub-design of $X$ locations closest
to $x$.  I.e., choose the $n$-{\em nearest neighbor (NN)
  sub-design} $X_n(x) \subseteq X$ and apply the predictive equations
(\ref{eq:predgp}--\ref{eq:preds2}) with $D_n(x) = (X_n, Y_n)$ where
$n$ is chosen as large as computational constraints allow.  This
is a sensible approach.  It is clearly an approximation: as
$n\rightarrow N$, $Y(x)|D_n$ converges to $Y(x)|D$ 
trivially.  Moreover, since the predictive equations are valid for any
design, interpreting the accuracy of the approximation involves the
same quantities as those obtained from an analysis based the on full
design.  Clearly, when $n \ll N$, the variance $V(x)|D_n$ can be
much larger than $V(x)|D$.  However, if $N$ is so big as to preclude
obtaining any predictor at all, within reasonable computational
constraints, obtaining one with marginally inflated
uncertainty is better than nothing.

Though simple, one naturally wonders if it is optimal for fixed $n$.  It is
not: \citet{vecchia:1988} observed that the best sub-design for prediction
depends $\theta$, and is usually not the NN design. In fact, NN is uniformly
suboptimal under certain conditions \citep{stein:chi:welty:2004}. Still,
finding the best design for $n > 1$ is hard, involving a high-dimensional
non-convex optimization.  Joint inference with $\theta$ can compound that
expense. Simple NN seems pragmatic by comparison, but we show how to do
better without much extra effort.

Our main idea, which is akin to forward stepwise variable selection in
regression, is as follows.  For a particular predictive location $x$, we
search for a sub-design $X_n(x)$, with accompanying responses $Y_n(x)$,
together comprising data $D_n(x)$, by making a sequence of greedy decisions
$D_{j+1}(x) = D_j(x) \cup (x_{j+1}, y(x_{j+1}))$, $j=n_0, \dots, n$.  Each
choice of $x_{j+1}$ depends on the previous choices saved in $D_j(x)$ by
searching over a criterion. Evaluating the criterion, and updating the
quantities required for the next iteration, must not exceed $O(j^2)$ so that
the total scheme remains $O(n^3)$, like NN.  We start with a very small
$D_{n_0}(x)$ comprised of $n_0$ NNs, which is easy to motivate in light of
results shown in Section
\ref{sec:illustrate}.


\subsection{Greedy search to minimize mean-squared predictive error}
\label{sec:greedy}

Given $x$ and $D_j(x)$, we search for $x_{j+1}$ by considering its
impact on the variance of $Y(x)$, taking into account
uncertainty in $\theta$, through an empirical Bayes mean-square
prediction error (MSPE) criterion: $ J(x_{j+1}, x) = \bE\{ [Y(x) -
\mu_{j+1}(x; \hat{\theta}_{j+1})]^2 | D_j(x) \}$, which should be
minimized. The predictive mean $\mu_{j+1}(x; \hat{\theta}_{j+1})$
follows (\ref{eq:predgp}), although we embellish notation here to
explicitly indicate dependence on $x_{j+1}$ and the future
$y_{j+1}$.  This mean represents the hypothetical prediction of $Y(x)$
arising after choosing $x_{j+1}$, observing $y_{j+1}$, and calculating
the MLE $\hat{\theta}_{j+1}$ based on the resulting $D_{j+1}(x)$.
Where convenient we drop the $x$ argument in the local design, $D_{j}
\equiv D_j(x)$.  Also note that $\hat{\theta}_j \equiv
\hat{\theta}_j(x)$ since it depends on $D_j(x)$.

In Section \ref{sec:mspe} we derive the approximation
\begin{equation}
J(x_{j+1},x) \approx  V_j(x | x_{j+1}; \hat{\theta}_j) + \left(\frac{\partial \mu_j(x;
    \theta)}{\partial \theta}
\Big{|}_{\theta = \hat{\theta}_j}\right)^2 /
\mathcal{G}_{j+1}(\hat{\theta}_j).
\label{eq:mspe}
\end{equation}
The terms in (\ref{eq:mspe}) are explained below in turn.
\begin{itemize}
\item $V_j(x|x_{j+1}; \theta)$ is the new variance 
  after adding $x_{j+1}$ into $D_j$, treating $\theta$ as
  known.
\begin{align}
V_j(x|x_{j+1}; \theta) &= \frac{\psi_j}{j-2} v_{j+1}(x; \theta),
\nonumber \\ 
\mbox{where } \;\;\; v_{j+1}(x; \theta) &= \left[ K_{j+1}(x, x) -
k_{j+1}^\top(x) K_{j+1}^{-1} k_{j+1}(x) \right]
\label{eq:newv}
\end{align}
captures $x_{j+1}$'s contribution to the reduction in variance by
placing its correlations to $X_j$ in the $j+1^\mathrm{st}$ row/col of
$K_{j+1}$.  The correlation between $x$ and $X_{j+1}(x) \equiv
(X_j(x), x_{j+1})$ comes in through the $j+1^\mathrm{st}$ entry of
$k_{j+1}$, the rest being $k_j$.  

\item $\frac{\partial \mu_j(x; \theta)}{\partial \theta}$ is the
  derivative of the predictive mean (\ref{eq:predgp}) at $x$,
  given $D_j$, with respect to $\theta$. 
\begin{equation}
\frac{\partial \mu_j(x;\theta)}{\partial \theta} = 
[\dot{k}_j(x) - \dot{\mathcal{K}}_j k_j(x)]^\top K_j^{-1} Y_j,
\end{equation}
where $\dot{\mathcal{K}}_j = \dot{K}_j K_j^{-1}$ and $\dot{k}_j(x)$ is the
$j$-length column vector of derivatives of the correlation function $K(x,
x_k)$, $k=1,\dots,j$, with respect to $\theta$.
\item $\mathcal{G}_{j+1}(\theta)$ is the expected Fisher
  information
     comprised of the information from $D_j$ plus the
  expected component from the future $y_{j+1}$ at $x_{j+1}$:
\begin{align}
  \mathcal{G}_{j+1}(\theta) &= F_j(\theta) + \bE \left\{-
    \frac{\partial^2 \ell_j(y_{j+1}; \theta)}{\partial \theta^2}
    \Big{|} D_j; \theta
  \right\}  \label{eq:G} \\
  &\approx F_j(\theta) + \frac{1}{2 V_j(x_{j+1};\theta)^2} \times
  \left( \frac{\partial V_j(x_{j+1}; \theta)}{\partial \theta}
  \right)^2 + \frac{1}{V_j(x_{j+1}; \theta)} \left( \frac{\partial
      \mu_j(x_{j+1}; \theta)}{\partial \theta} \right)^2, \nonumber
\end{align}
where $F_j(\theta) = - \ell''(Y_j; \theta)$ from (\ref{eq:d2l}),
$V_j(x_{j+1}; \theta) \equiv V(x_{j+1}| D_j; \theta)$ following
(\ref{eq:preds2}), and
\begin{align*}
 \frac{\partial V_j(x; \theta)}{\partial \theta}  & =
 \frac{Y_j^\top K_j^{-1}\dot{\mathcal{K}}_j Y_j }{j-2} 
 \left(K(x, x) -k_j^\top(x)
  \tilde{k}_j(x)\right) \\
&\;\;  -\!\frac{\psi_j}{j-2}\left[ \dot{k}_j^\top(x) \tilde{k}_j(x) + \tilde{k}_j(x)^\top 
(\dot{k}_j(x)\!-\!\dot{\mathcal{K}}_j k_j(x)) \right]\!,
\mbox{ where } \tilde{k}_j(x) = K_j^{-1} k_j(x).
\end{align*}
\end{itemize}

The approximation in (\ref{eq:mspe}) is three-fold: (a) the partial derivative
of $\mu_{j+1}$ is approximated by that of $\mu_j$; (b) the Student-$t$
predictive equations are approximated by moment-matched Gaussian ones
(\ref{eq:G}); and (c) the future $\hat{\theta}_{j+1}$ is replaced by the
current estimate $\hat{\theta}_j$ based on the data $D_j$ obtained so far.
All are required since averaging over future {\em expected} $y_{j+1}$ is
analytically intractable.  Even though a $y_{j+1}$ is technically available in
our context, we deliberately do {\em not} condition on this value, which would
result in the undesirable avoidance of selecting an $x_{j+1}$ whose $y_{j+1}$
is at odds with $D_j$.  [More discussion in \ref{sec:mspe}.]

Observe that when the Fisher information is large, the MSPE
(\ref{eq:mspe}) is dominated by the reduced variance for $x_{j+1}$,
the first term.  We return to this special case in Section
\ref{sec:special}.  For smaller $j$ the extra term gives preference to
choosing an $x_{j+1}$ that is expected to improve estimates of
$\theta$ nearby to $x$.  Considering that a sensible non-local design
strategy is to maximize (the determinant of) the Fisher information,
the MSPE is thus making a local adjustment by (a) weighting the Fisher
information by the rate of change of the predictive mean near $x$ and
(b) balancing that with the aim of reducing local predictive variance.
The appropriate balance is automatically dictated by the definition of
$J(x_{j+1}, x)$, which considers the effect of uncertainty in $\theta$
on the predictive uncertainty in $Y(x)$.  Aspect (a) is suggestive of
potential benefit in estimating a localized, i.e., nonstationary,
spatial field.

\subsection{Fast updates}
\label{sec:update}

All required quantities can be 
updated quickly, from iteration $j$ to $j+1$,
in $O(j^2)$ time so that
$n$ applications are in $O(n^3)$. This is possible since
$K_{j+1}^{-1}$ can be obtained efficiently from $K_j^{-1}$, both
conditional on the same parameter $\theta$, via the partition inverse
equations [see \ref{sec:pie}].  
 We use them in several applications.
For example,
\begin{align}
 v_j(x&; \theta)  - v_{j+1}(x; \theta), \quad \mbox{(dropping $\theta$ below
for compactness)} \label{eq:dxy} \\ &= k_j^\top(x) G_j(x_{j+1}) v_j(x_{j+1})
k_j(x) + 2k_j^\top(x) g_j(x_{j+1}) K(x_{j+1},x) + K(x_{j+1},x)^2 /
v_j(x_{j+1}), \nonumber
\end{align}
where $G_j(x') \equiv g_j(x') g_j^\top(x')$, and $g_j(x') = K_j^{-1}
k_j(x')/v_j(x')$.
Each product above requires most $O(j^2)$ time. Adding a row and
column $k_j(x_{j+1})$ to $K_j$ 
yields $K_{j+1}$ in $O(j)$ extra time.
%
The log marginal likelihood requires $\log |K_j|$, which may similarly be updated
as
\begin{align*}
\log |K_{j+1}| &= \log |K_j| + \log(K(x_{j+1}, x_{j+1}) + g_j^\top(x_{j+1})
k_j(x_{j+1})v_j(x_{j+1})) \\ &= \log |K_j| + \log(v_j(x_{j+1})).
\end{align*}
in $O(j)$ time.
Fast updates are
 available for key aspects of the predictive equations
(\ref{eq:predgp}--\ref{eq:preds2}):
\begin{align}
K_{j+1}^{-1} Y_{j+1} &= 
\begin{pmatrix} K_j^{-1} Y_j + g_j(x_{j+1})(
h_j(x_{j+1})v_j(x_{j+1}) + y_{j+1}) \\ 
h_j(x_{j+1}) + y_{j+1}/v_j(x_{j+1})
\end{pmatrix} \nonumber \\
\mbox{and } \;\; \psi_{j+1} &= \psi_j + h_j(x_{j+1})^2 v_j(x_{j+1})
  + 2 y_{j+1} h_j(x_{j+1}) + y_{j+1}^2 / v_j(x_{j+1}), \label{eq:psiup}
\end{align}
where $h_j(x_{j+1}) = Y_j^\top g_j(x_{j+1})$.
Both updates take time in $O(j)$.

For fast updates of the observed information $F_{j+1}(\theta)$, write
the log likelihood as a sum of components from $D_{j}$ and from
$(x_{j+1}, y_{j+1})$, with the latter denoted by $\ell_j(y_{j+1};
\theta)$.  This suggests the simple recursion $F_{j+1}(\theta) =
F_{j}(\theta) + \frac{\partial^2 \ell_j(y_{j+1}; \theta)}{\partial
  \theta^2}$.  In \ref{sec:mspe} we show that
\[
\frac{\partial^2 \ell_j(y_{j+1}; \theta)}{\partial
  \theta^2} = - \frac{\ddot{V}_j}{2 V_j} \!+\! \frac{\dot{V}_j^2}{2
  V_j^2} \!-\! \frac{\dot{\mu}_j^2}{V_j}
  + (y_{j+1} - \mu_j)\!\left(\frac{\ddot{\mu}_j}{V_j} - 2 \frac{\dot{\mu}_j
  \dot{V}_j}{V_j^2}\right) 
 + (y_{j+1} - \mu_j)^2\! \left( \frac{\ddot{V}_j}{2 V_j^2} -
   \frac{\dot{V}_j^2}{V_j^3}
\right).
\]
For compactness we are suppressing arguments to the mean and variance
functions, and shorthanding their derivatives.  E.g., $V_j \equiv
V_j(x_{j+1}; \theta)$ and $\ddot{\mu}_j \equiv \frac{\partial^2
  \mu_j(x_{j+1}; \theta)}{\partial \theta^2}$.  Expressions for second
derivatives of predictive quantities may als be found in \ref{sec:mspe}.
Observe that the expectation (given $D_j$) of the above quantity
augments $F_j(\theta)$ to complete $\mathcal{G}_{j+1}(\theta)$ in
(\ref{eq:G}).

Updates of all other quantities required for the calculations in Section
\ref{sec:greedy} follow trivially.  Since they require fixed $\theta$, our
greedy scheme is only efficient if MLEs are {\em not} recalculated after each
sequential design decision, suggesting further approximation is warranted to
obtain a thrifty MSPE-based local design.  We find that working with a
reasonable fixed value of $\theta$ throughout the iterations works well.  At
the end of the $n$ local design and update steps we can find the MLE
$\hat{\theta}_n(x)$ by Newton-like methods with analytic derivatives
(\ref{eq:dl}--\ref{eq:d2l}), incurring an $O(n^3)$ cost once. More
details on local inference are deferred to Section
\ref{sec:param}.

\subsection{A special case}
\label{sec:special}

When the Fisher information is large, the MSPE (\ref{eq:mspe}) reduces to
$V_j(x| x_{j+1}; \theta)$, which is the new variance at $x$ when $x_{j+1}$ is
added into the design [Section
\ref{sec:greedy}].  A similar statistic has been used in the
sequential design of computer experiments before, but in a different
context: choosing design locations for new computer
simulations \citep[e.g.,][]{seo:etal:2000,gra:lee:2009}.  Rather than
focus on a single $x$, those works average
reductions in variance over the whole input space $\mathcal{X}$:
$\Delta V_j(x_{j+1}; \theta) = \int_\mathcal{X} (V_j(x;\theta) - V_j(x|
x_{j+1};\theta)) \, dx. \label{eq:alc} $ Part of the integrand does not
depend on $x_{j+1}$, and so can be ignored.  The other part is
proportional to $v_{j+1}(x; \theta)$ in (\ref{eq:dxy}).  In the
special case where $K(\cdot, \cdot)$ yields the identity
\citep{anag:gra:2012}, or when $\mathcal{X}$ is a finite and discrete
set, the integral is also analytic.  Otherwise numerical methods are
needed, which may add significant computational burden.

Choosing $x_{j+1}$ to maximize $\Delta V_j(x_{j+1};\theta)$ is a sensible
heuristic since it augments the design with an input-output pair that has the
most potential to reduce variance globally.  In repeated application,
$j=1,\dots, N$, the resulting designs have been shown to approximate maximum
information designs \citep{cohn:1996}. To credit its originator, subsequent
authors have referred to this technique as {\em active learning Cohn} (ALC).
However, the expensive numerical integration required means that
alternative heuristics are usually preferred.

A local design criteria based on the integrand alone---the simplest
implementation of which involves maximizing (\ref{eq:dxy})---may therefore be
a sensible alternative to MSPE, not just further approximation. It has
information theoretic motivation and shortcuts cumbersome derivative
calculations. In what follows we overload the terminology a bit and refer to
local design based on maximizing (\ref{eq:dxy}) as ALC. However, it is
important to note that its application in this context is novel.  ALC in its
original form has never been used to select a local sub-design and, since no
integration is needed (numerical or otherwise) the computation is
straightforward.  In \ref{sec:pcor} we show that this heuristic can be
derived by studying partial correlations akin to those used to obtain
sequential least squares estimators.

\subsection{An illustration}
\label{sec:illustrate}

Consider a surface first studied by
\citet{gramacy:lee:2011}, defined by
$f(x_1,x_2) = -w(x_1)w(x_2)$, where
$w(x) = \exp\left(-(x-1)^2\right) + \exp\left(-0.8(x+1)^2\right) 
- 0.05\sin\left(8(x+0.1)\right)$,
generating a data set of $x$-$y$ pairs on dense $201\times 201$
($=40401$ point) regular grid in $[-2,2]$.  
\begin{figure}[ht!]
\centering
\includegraphics[scale=0.4]{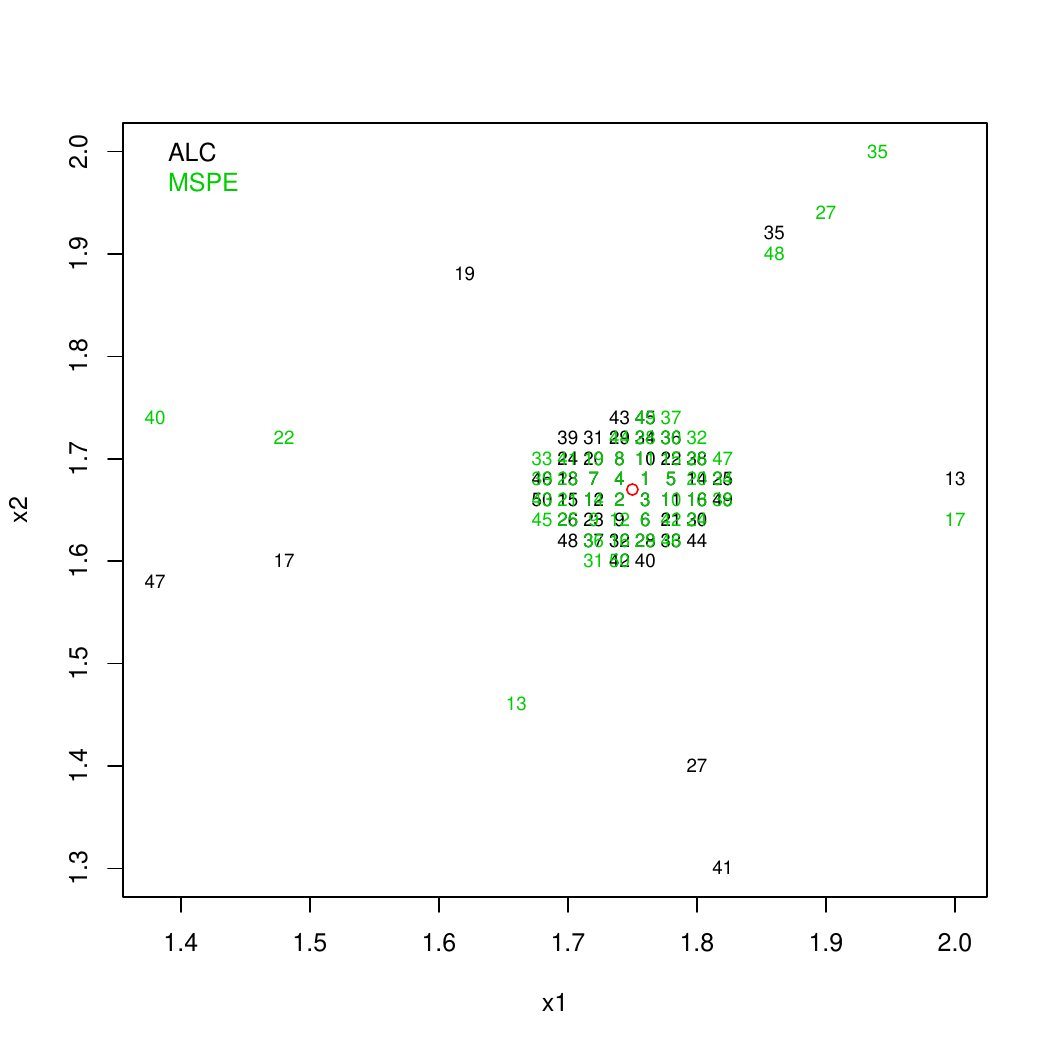} \hspace{1cm}
\includegraphics[scale=0.4]{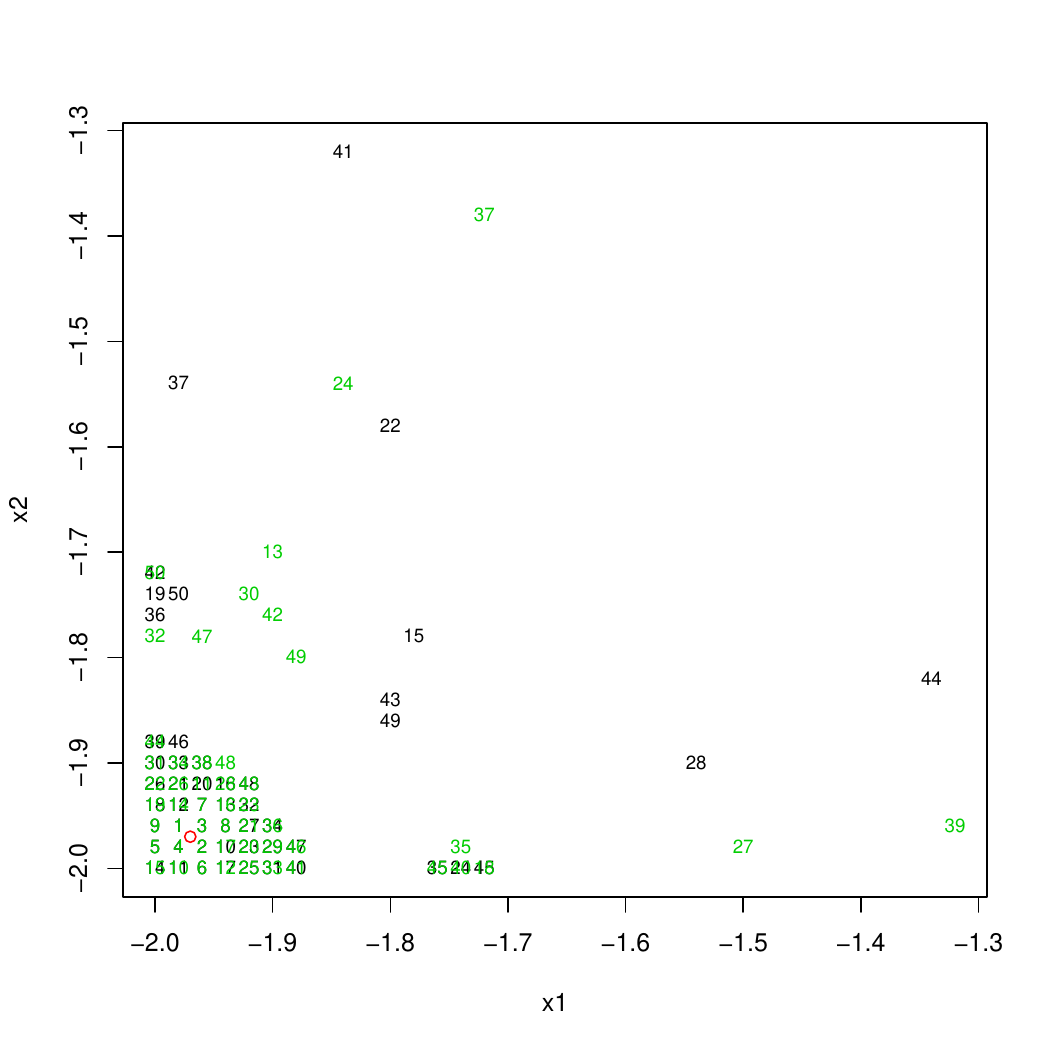}
\caption{A comparison between ALC (black) and MSPE (green) heuristics at two
  predictive locations. The numbers plotted indicate the order in which
  the design sites were chosen.}
\label{f:alcmspe}
\end{figure}
Figure \ref{f:alcmspe} shows local designs for two input locations, $x$, under
MSPE and ALC criteria.  Both share the same initial design $D_{n_0}$
comprising the closest six NNs to $x$.  Although the two heuristics agree on
some sites chosen, observe that they are not selected in the same order after
iteration nine.  The patterns diverge most for further out sites, but their
qualitative flavor remains similar.  In both cases, the bounding box
$[-2,2]^2$ impacts local design symmetry.

The full 46 local design iterations take less than a tenth of a second
on a modern iMac, although in repeated trials (to remove OS noise) we
found that MSPE takes 2-3 times longer. Inverting a $40K \times 40K$
matrix cannot, to our knowledge, be done on a modern desktop primarily
due to memory swapping issues.  For a point of reference, inverting a
$4000 \times 4000$ matrix took us about seven seconds using a
multithreaded BLAS/Lapack.  Memory issues aside we deduce that the
alternative full-GP result would have taken days.

\begin{figure}[ht!]
\centering
\includegraphics[scale=0.55,trim=10 10 5 10]{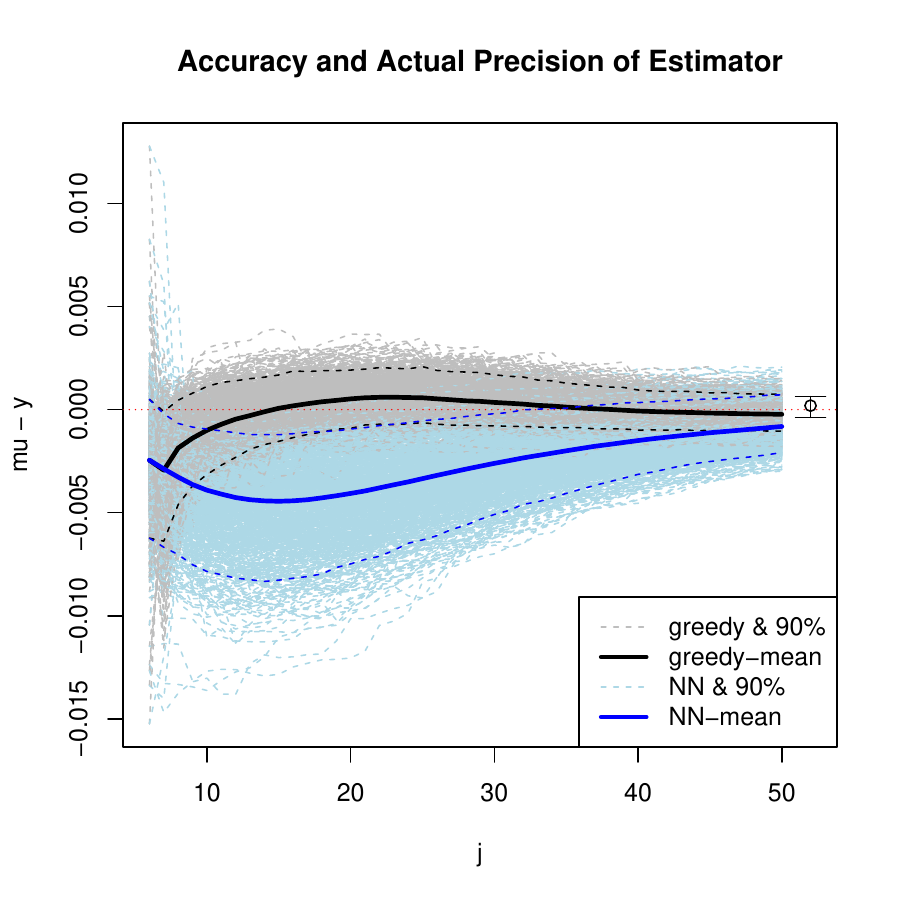}\hfill
\includegraphics[scale=0.55, trim=10 10 10 10]{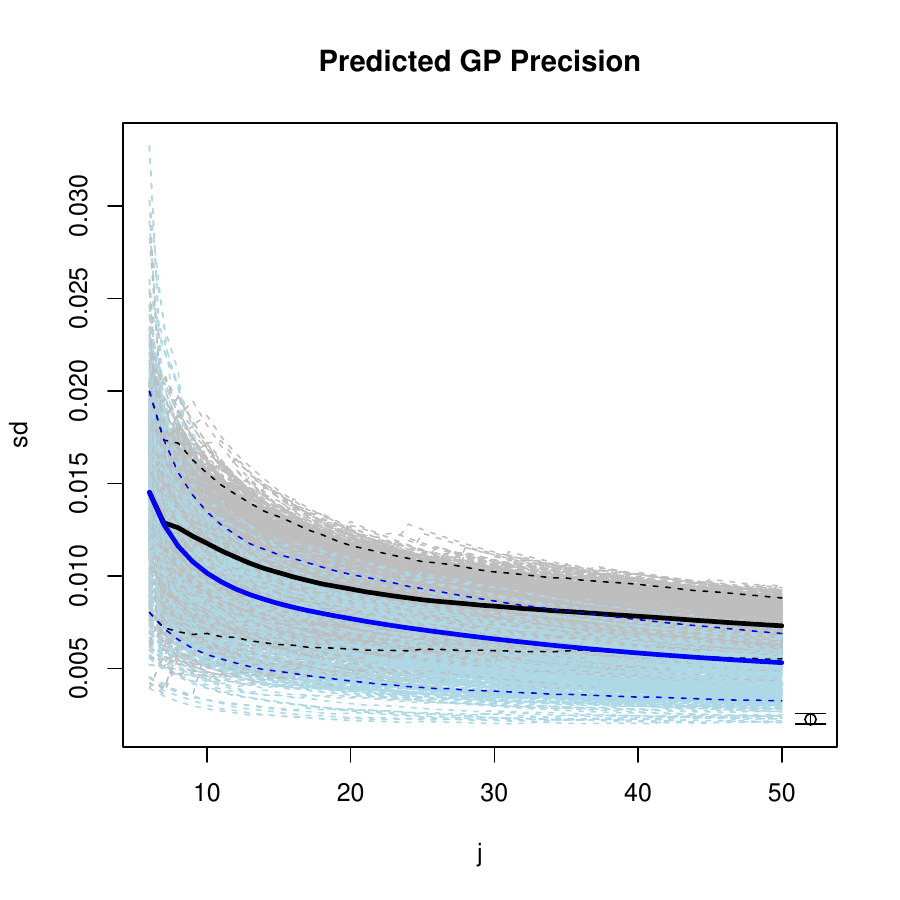}
\caption{Comparing the NN predictor to the greedy  approach
  in 1000 repeated repetitions an LHS of size 10001: 10K train/1 test.
  Both methods are initialized with an NN design $n_0 = 6$. The {\em
    left} pane shows the actual bias and precision and the {\em right}
  shows the predicted precision (GP
  predictive standard deviation).  Each repetition is shown in light shades; the
  darker dashed lines are 90\% quantiles, and the dark solid line is
  the average.  The open-circles with whiskers at $j=52$ are the 
  90\% quantile and average results from a size 1000 NN design.}
\label{f:compare}
\end{figure}

Figure \ref{f:compare} illustrates the accuracy of the approximate predictive
distribution thus obtained, with comparison to the NN alternative.  These
results summarize the output of a Monte Carlo experiment repeated 1000 times,
whose setup is similar to that above with the following exception.  In each
repetition we generated a Latin Hypercube sample (LHS)
of size 10001, and treated the
first 10K as training $(X,Y)$ and the last one as our testing $(x,y)$.  We
then obtained the predictive equations under NN and greedy/MSPE local designs
(to $x$) of size $n=50$, where both share a starting NN design $D_{n_0}$ of
size $n_0 = 6$.  Each lightly-shaded line in the plot represents one Monte
Carlo iteration.  The dark-dashed lines outline point-wise 90\% intervals, and
the dark-solid lines show the point-wise averages.  The open circles and
whiskers at 52 on the $j$-axis provide an approximate benchmark, based on a NN
local design of size 1000, for the best possible results.

The {\em left} pane shows the bias, defined as the predictive mean
$\mu_n(x)$ minus the true $y$-value.  Notice how, relative to the
greedy/MSPE method, the NN approach is slow to converge to the true
value and is consistently biased low after the initial design.  Both
start biased-low, which we attribute to $0$-mean reversion typical 
(for GPs) in areas of the input space sparsely covered by the design.
The {\em right} pane shows the predicted standard deviation over
design iterations.  Notice that the NN standard deviations are
consistently lower than those from the greedy/MSPE method.  So the NN
method is both biased-low {\em and} more confident than its comparator.
At $j=n=50$ both methods obtain a predictive mean close to
the truth, with lower variability for the greedy/MSPE
method.  Moreover, both adequately acknowledge uncertainty inherent in
the approximation obtained using a much smaller local design compared
to one obtained with orders of magnitude larger designs.  In the
\ref{sec:condum} we show that an added benefit of the greedy
approach is that the condition numbers of $K_j$ are lower than those
obtained from NN.

\section{Global emulation on a dense design}
\label{sec:global}

The simplest way to extend the analysis to cover a dense design of predictive
locations $x\in \mathcal{X}$ is to serialize: loop over each $x$ collecting
approximate predictive equations, each in $O(n^3)$ time.  For $T =
|\mathcal{X}|$ the total computational time is in $O(Tn^3)$.  Obtaining each
of the full GP sets of predictive equations, by contrast, would require
computational time in $O(T N^2 + N^3)$, where the latter $N^3$ is attributable
to obtaining $K^{-1}$.  One of the nice features of standard GP emulation is
that once $K^{-1}$ has been obtained the computations are fast $O(N^2)$
operations for each location $x$.  However, as long as $n \ll N$ our
approximate method is even faster despite having to rebuild and re-decompose
$K_j(x)$'s for each $x$.

\subsection{Parallelization}

The approximation at $x$ is built up sequentially, but completely
independently of other predictive locations.  Since a high degree of local
spatial correlation is a key modeling assumption this may seem like an
inefficient use of computational resources, and indeed it would be in serial
computation for each $x$. However, independence allows trivial
parallelization.  Predicting at a dense design of 10K locations with 8 threads
on an 4-core hyperthreaded (i.e., behaving like 8 cores) iMac, with a 40K design [as in Section
\ref{sec:illustrate}] took about 30 seconds and required only token
programmer effort: we used an {\tt OpenMP} pragma to parallelize a
single ``{\tt for}'' loop over $x
\in \mathcal{X}$ giving nearly linear speedup.  See the {\tt aGP} function in
the {\tt laGP} package.

\subsubsection*{Illustration continued}

\begin{figure}[ht!]
\centering
\begin{minipage}{5.5cm}
\includegraphics[scale=0.56, trim=30 80 20 70]{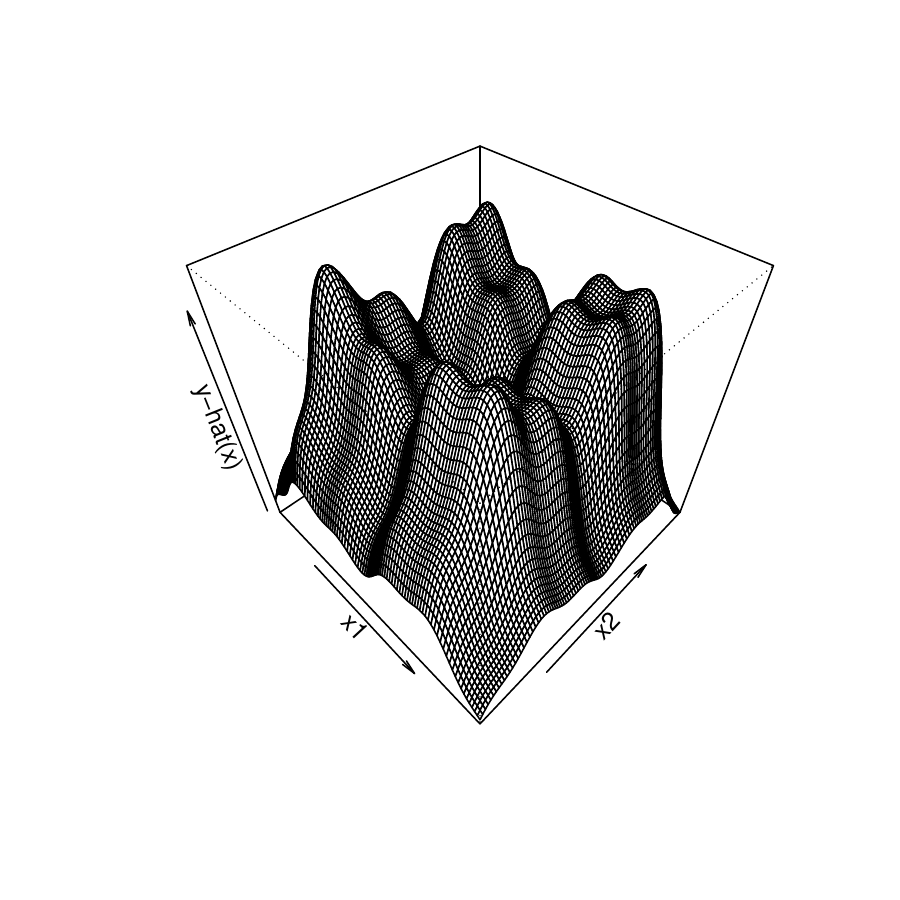}
\end{minipage}
\hspace{2cm}
\begin{minipage}{7cm}
\includegraphics[scale=0.6, trim=10 10 10 50]{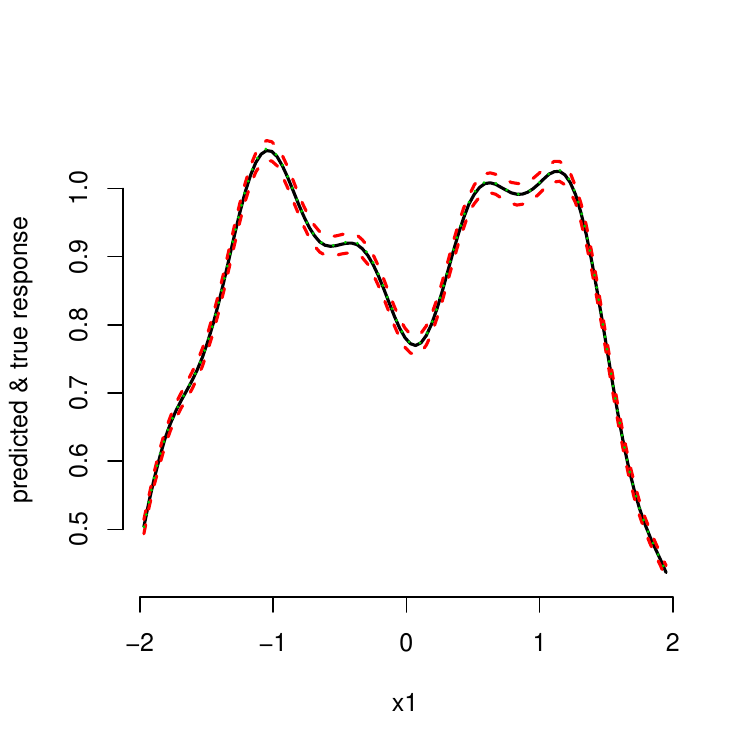}
\end{minipage}
\begin{minipage}{5.5cm}
\includegraphics[scale=0.6, trim=10 10 10 50]{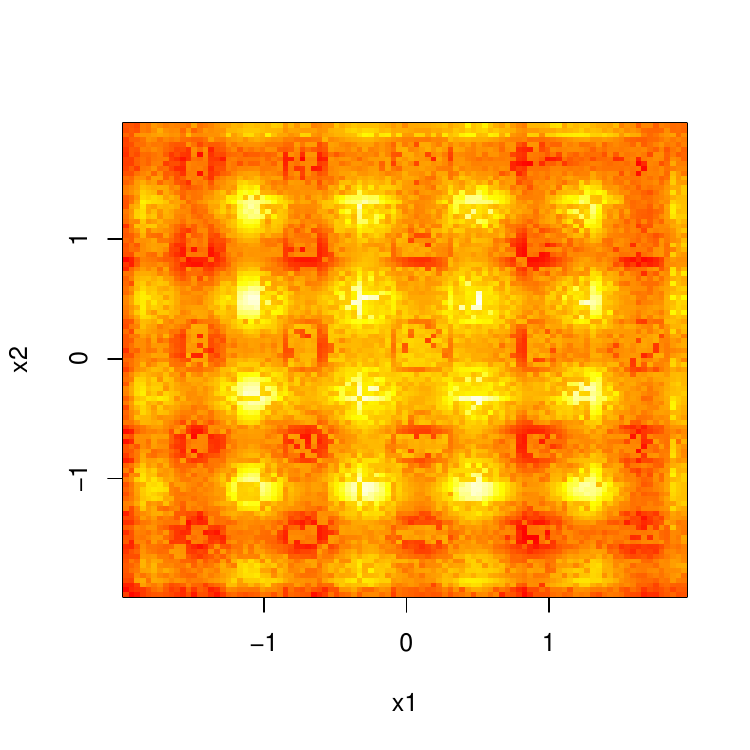}
\end{minipage}
\hspace{2cm}
\begin{minipage}{7cm}
\includegraphics[scale=0.6, trim=10 10 10 50]{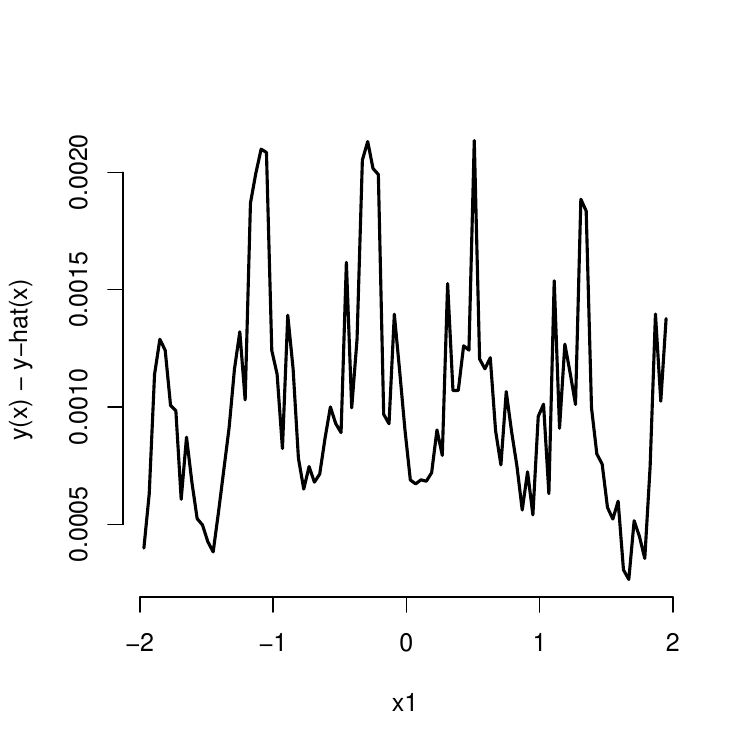} 
\end{minipage}
\begin{minipage}{5.5cm}
\includegraphics[scale=0.6, trim=10 10 10 50]{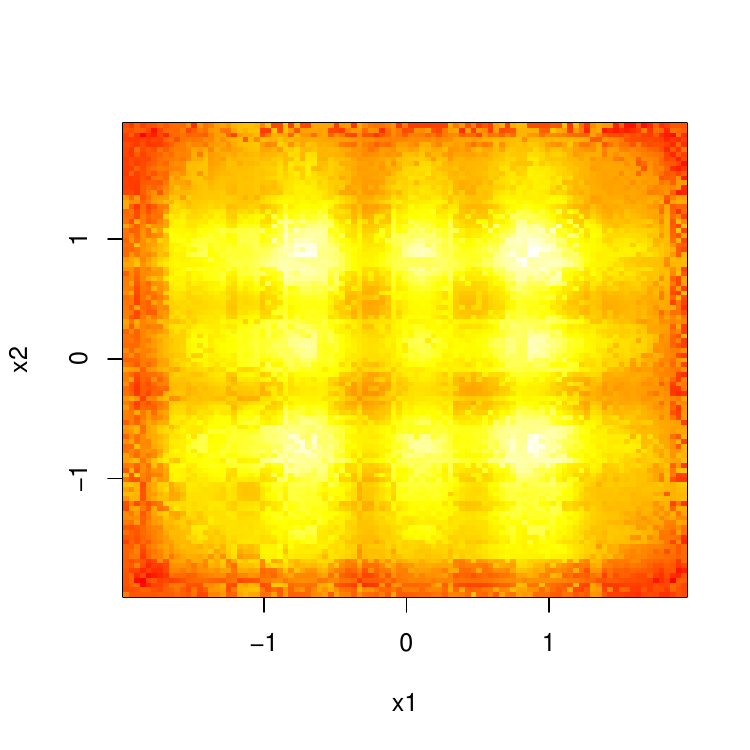}
\end{minipage}
\hspace{2cm}
\begin{minipage}{7cm}
\includegraphics[scale=0.6, trim=10 10 10 50]{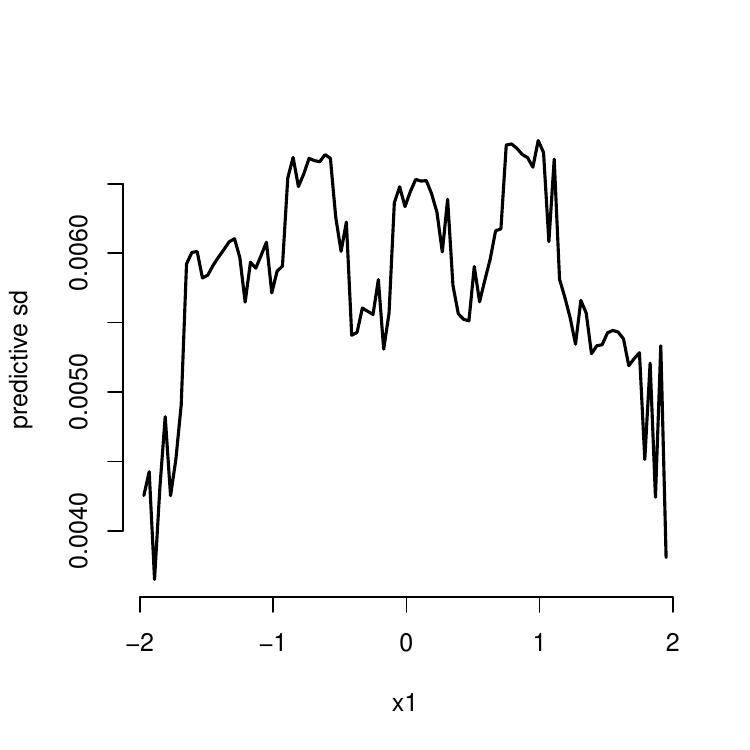} 
\end{minipage}
\caption{The {\em left} column shows 2d visualizations of the
  posterior mean surface $\mu(x|D_n)$, difference between the mean and the
  truth $\mu(x|D_n) - y(x)$ and the predicted standard deviation
  $\sqrt{V(x|D_n)}$.  In the image plots, lighter colors (yellows/whites) are
  higher values; darker (reds) are lower. The {\em right} column shows the
  same for the slice $x_2 = 0.5$.}
  \vspace{-1cm}
\label{f:global}
\end{figure}

Figure \ref{f:global} offers visualizations of the behavior of our
greedy, local, approximations applied globally in the input space.
The {\em left} column contains surfaces for both input dimensions,
whereas the {\em right} one shows a 1d slice for $x_2 = 0.5$.  The mean
surfaces ({\em top} row) exhibit essentially no visual discrepancy to
the truth.  There are, however, small discrepancies ({\em middle} row)
and the level of accuracy is not uniform across the input space.
Pockets of higher (though still small) inaccuracy are visible.
Finally, despite the uniform gridded design, the predicted standard
deviation ({\em bottom} row) is non-uniform throughout the input
space, appearing to smoothly vary albeit across a narrow range of
values.  This subtly nonstationary behavior is due to the localized
estimation of $\tau^2$ via $\psi(x)$.  It is not due to $\theta$
since that was fixed globally.  Closer inspection reveals that areas
of lower predictive variability ({\em bottom} row) correspond to
greater inaccuracy ({\em middle}).  Both results suggest that the
localized greedy approximation struggles to cope with the assumed
stationary covariance, and may benefit from a more
deliberate localized approach to the estimation of spatial
correlation.  

\subsection{Localized parameter inference and stagewise design}
\label{sec:param}


Wishing to leverage fast sequential updating, retain independence in
calculations for each predictive location $x$ so they can be
parallelized, and allow for local inference of the correlation
structure for nonstationary modeling, we propose the following
iterative scheme.

{
\begin{enumerate}
\item Choose a sensible starting global lengthscale parameter
  $\theta_x = \theta_0$ for all $x$.
\item Calculate local designs $D_n(x, \theta_x)$ based on sequential
  application of the MSPE or ALC design heuristics, 
  independently for each $x$.
\item Also independently, calculate the MLE lengthscale
  $\hat{\theta}_n(x) | D_n(x, \theta_x)$ thereby explicitly obtaining a
  globally nonstationary predictive surface.
\item Set $\theta_x = \hat{\theta}_n(x)$ possibly after smoothing
  spatially over all $x$ locations.
\item Repeat steps 2--4 as desired.  Then output predictions each $x$, 
independently, based on $D_n(x)$ and possibly smoothed $\theta_x$.
\end{enumerate}
}
Usually only one repetition of steps 2--4 is required for joint convergence of
local designs $D_n(x)$ and parameter estimates $\hat{\theta}_n(x)$, for all
locations $x$.  The initial $\theta_0$ is worthy of some consideration, since
certain pathological values---very small or very large on the scale of squared
distances in the input space---can slow convergence.  We find that a sensible
default $\theta_0$ setting can be chosen from the lower quantiles of squared
distances in $X$.

MLE calculations in Step 3 proceed via Newton-like methods as described in
Section \ref{sec:update}.  We take $\theta_x$ for initialization, which may be
$\theta_0$ in the first stage or a possibly smoothed $\hat{\theta}_n(x)$ in
later stages. Convergence is very fast in the latter case (1--4 iterations) as
we illustrate below.  The former can require about twice as many iterations
depending on $\theta_0$, however this can be reduced if nearby
$\hat{\theta}_n(x')$ can be used for initialization instead.  As a share of
the computational cost of the entire local design scheme, the burden of
finding local MLEs is dwarfed by the MSPE calculations when $N\gg n$.  So
reducing the number of Newton steps may not be a primary concern in practice.

The smoothing suggestion in Step 4 stems from both pragmatic and modeling
considerations.  GP likelihood surfaces can be multimodal so the Newton method
is not guaranteed to find a global mode.  Smoothing can offer some protection
against rapid mode-switching in local estimators.  On the modeling end, it is
sensible to posit {\em a priori} that the correlation parameters be
constrained somewhat to vary slowly across the input space.  
However, we find that this step is
not absolutely necessary to obtain good prediction results, though it may aid
in convergence of the two-stage scheme.

\subsubsection*{Illustration continued}

\begin{figure}[ht!]
\includegraphics[scale=0.34,trim=30 0 30 0]{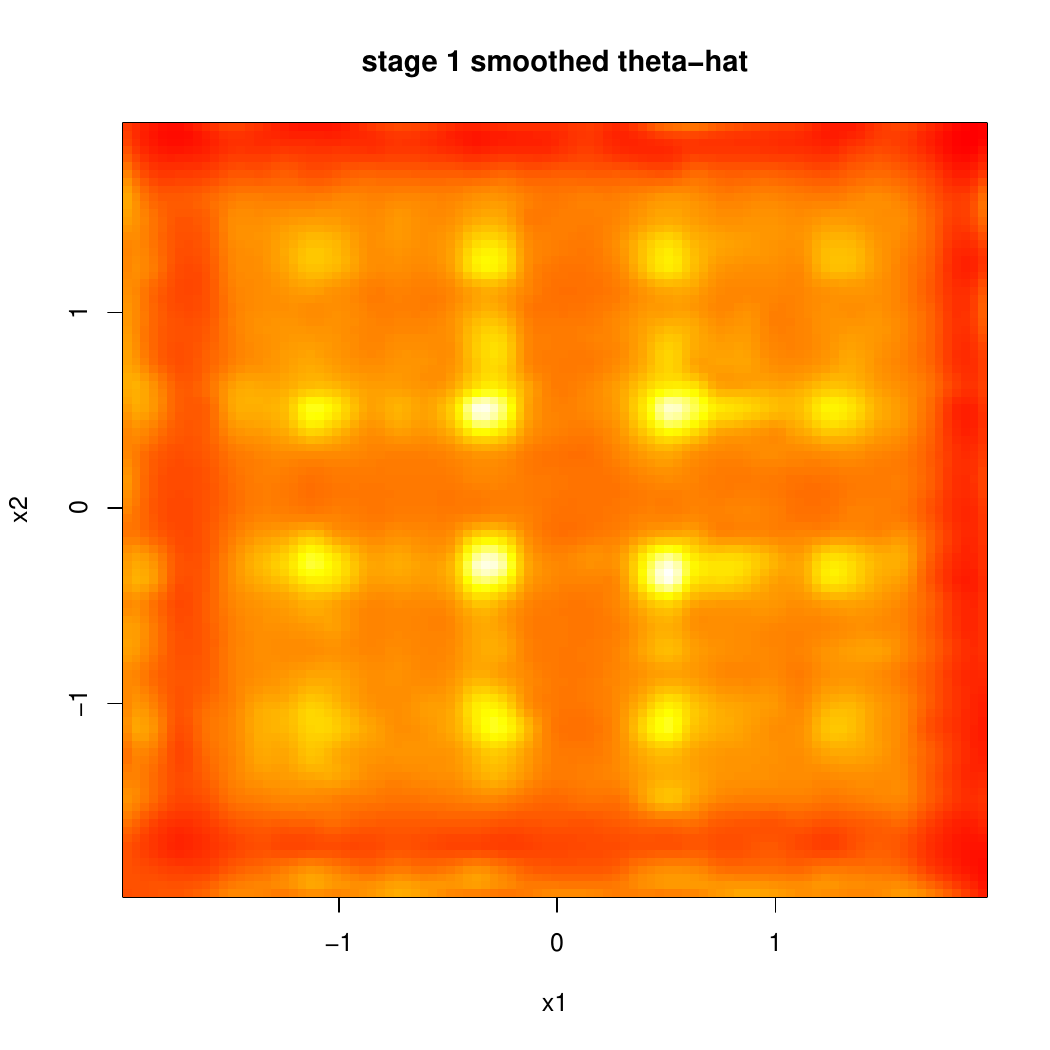} \hfill
\includegraphics[scale=0.34,trim=59 0 0 0,clip=TRUE]{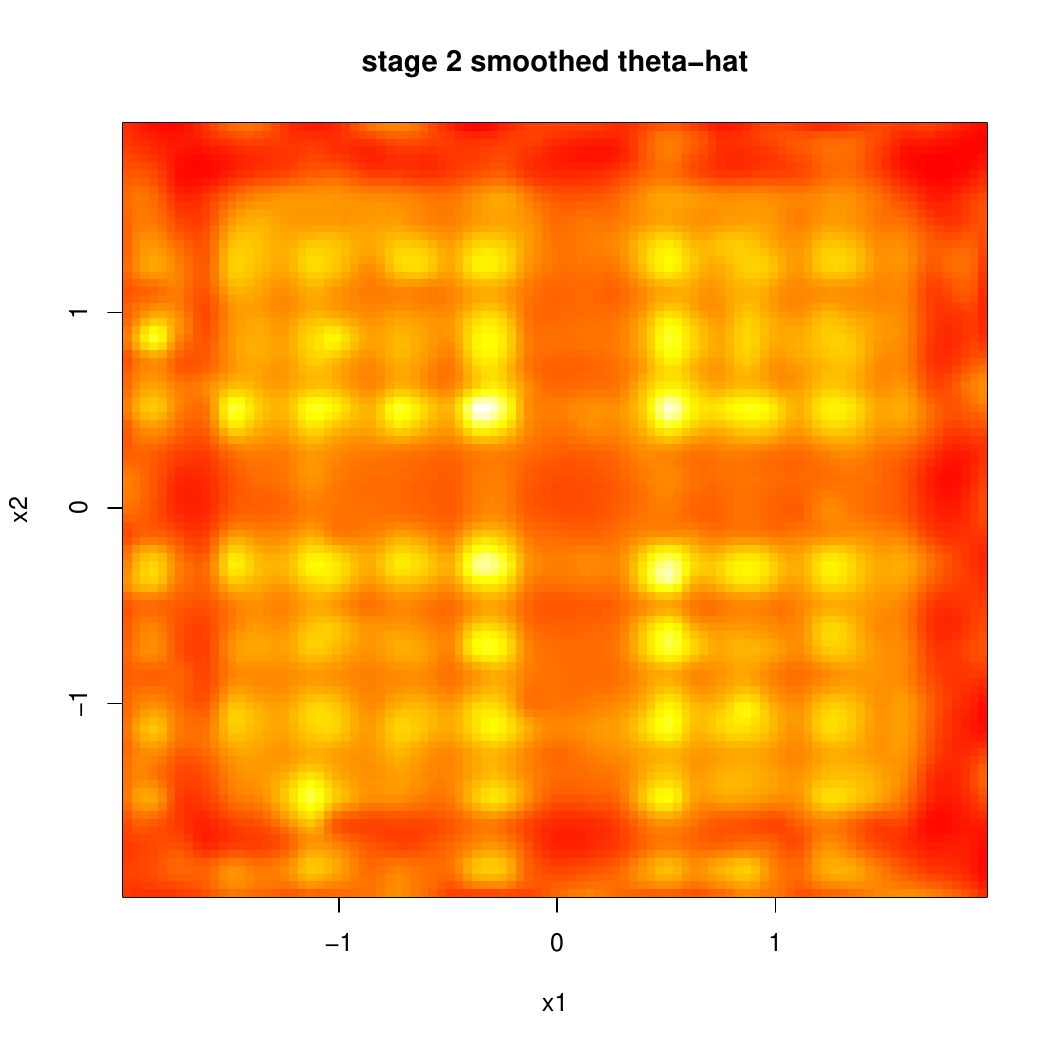} \hfill
\includegraphics[scale=0.34,trim=40 0 30 0]{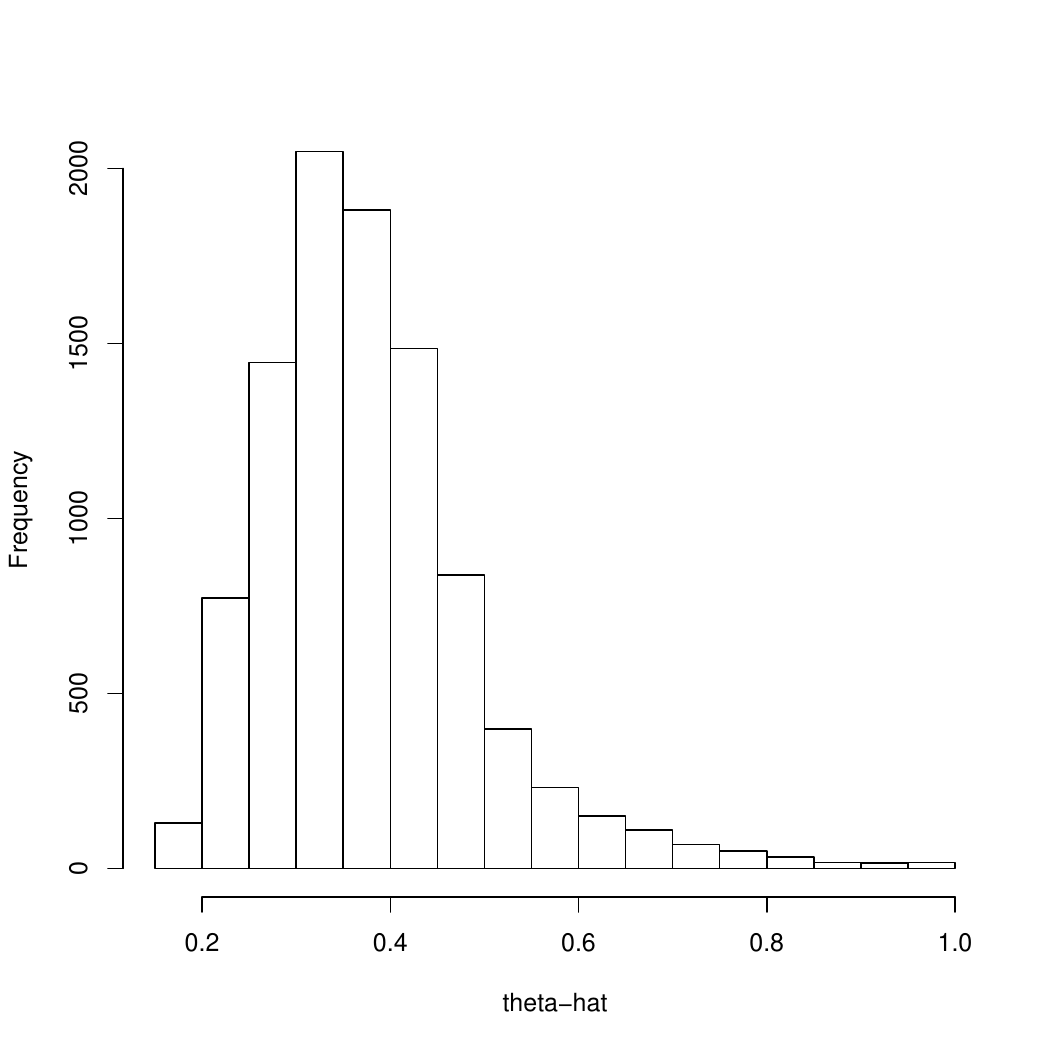}
\caption{Smoothed $\log \hat{\theta}_n(x)$ values, using  {\tt loess} in
{\sf R} with {\tt span=0.01}, after the first and
  second stages (reds are lower values), and by a histogram of
  values from the second stage.}
\label{f:dmle}
\end{figure}

Figure \ref{f:dmle} shows the estimated $\log$ local lengthscale parameters on
our illustrative example from Section \ref{sec:loc} using ALC; the
MSPE results are similar but require 50\% more computational effort.
The first pane shows a snapshot after the first stage; the second
after the second stage. 
Notice the increase in fidelity from one stage to the next, and 
the similarity to the predictive error and variance results from
Figure \ref{f:global}, with larger lengthscales corresponding to areas poorer
emulation under the model with a fixed, global, lengthscale.  
The final pane in Figure \ref{f:dmle} 
shows a histogram of the $\hat{\theta}_n(x)$'s, for $n=50$, obtained
after the second stage.  The Newton steps from the first stage required about
6.5 iterations on average, with 90\% between 6 and 8 whereas the second stage
took 4.2 (3 to 6) when primed with previous stage values.

\section{Empirical demonstration and comparison}
\label{sec:results}

\vspace{-0.25cm}
\subsection{Illustrative example}
\label{sec:rill}

We start by making an out-of-sample comparison of our local approximation
methods on the prediction problem described above. Comparators include
variations based on NN, and greedy selection via MPSE and ALC.  All greedy
methods are initialized with $\theta_0 = 0.7$ which is in the lower quartile
of squared distances in $X$. We furthermore consider a second iteration of MSPE and
ALC local design initialized with smoothed MLEs $\hat{\theta}_n(x)$ from the
first stage.  NN with $\hat{\theta}_n(x)$ is included too, which we regard as
one of our novel (if small) contributions. Note that fully placing NN in the
context of the scheme outlined in Section
\ref{sec:global} has limits since later stage local NN designs would be
identical to those from the first stage, as they do not depend on $\theta$. We
consider variations where MLEs are not calculated. In those cases we used
$\theta_0 = 1$ which is a reasonable but not optimal value, inside but towards
the upper extreme of the 95\% interval of $\hat{\theta}_n(x)$-values found,
{\em ex poste}, by the other methods.  The timings were obtained with 8
threads on $4$-core hyperthreaded iMac, via {\tt OpenMP} pragmas. All other
particulars are exactly as described by the illustrations above
We also consider a ``big'' NN
benchmark, which uses $n=200$.  The others used $n=50$, as previously.

Table \ref{t:compare} summarizes execution time in seconds, predictive
accuracy with RMSE, uncertainty with pointwise 95\% intervals, etc. All
methods exhibit high precision on the scale of the response, and all over
cover.  The best methods are three times better than the worst with, generally
speaking,  more computational effort reaping rewards.  Methods which do not
infer $\hat{\theta}_n(x)$ fare worst, and NN is only competitive with the
greedy methods when it uses an order of magnitude larger local designs
(``nnbig''). Although competitive in terms of accuracy per unit time, even
giving lower predicted standard deviation, this success comes at the expense
of storing a 4x larger output object, which may limit portability.

\subsection{Comparison to compactly supported covariances}
\label{sec:rcsc}

For a broader comparison, consider the borehole function \citep{worley:1987}
which is a common benchmark for computer experiments
\citep[e.g.][]{morris:mitchell:ylvisaker:1993}.  The response $y$ is given by
\begin{equation}
y = 2\pi T_u [H_u - H_l] \left/
\log\left(\frac{r}{r_w}\right) 
\left[1 + \frac{2 L T_u}{\log (r/r_w) r_w^2 K_w} + \frac{T_u}{T_l}
\right] \right.
\,.
\label{eq:borehole}
\end{equation}
The eight inputs are constrained to lie in a rectangular domain:
\begin{align*}
r_w &\in [0.05, 0.15] & r &\in [100,5000] & T_u &\in [63070, 115600] &
T_l &\in [63.1, 116] \\
H_u &\in [990, 1110] & H_l &\in [700, 820] & L &\in [1120, 1680] & 
K_w &\in [9855, 12045].
\end{align*}
Part of our reason for choosing this data is that it was used as an
illustrative example for the CSC method described by
\cite{kaufman:etal:2012}, which may be the most widely adopted modern
approach to fast approximate emulation for computer experiments.  The
authors provide a script illustrating their {\sf R} package {\tt
  sparseEM} on this data.
\begin{center}
\url{http://www.stat.berkeley.edu/~cgk/rcode/assets/SparseEmExample.R}
\end{center}
Here we follow that script
verbatim, except augmenting their 99\% sparse estimator with a 99.9\% one
for faster results.  
They generate a LHS of size $4500$, using the first 4000 for training and the
rest for testing, and use a statistic called NSE
for evaluation.  As we argue in \ref{sec:tabs}, raw NSE values can be hard to interpret when
measuring the accuracy of deterministic computer experiments, so we
instead report report $\sqrt{1-\mathrm{NSE}} = \mathrm{sd}(\mbox{prediction
error})/\mathrm{sd}(\mbox{response})$, and in addition to other
metrics like RMSE. 

\begin{table}[ht!]
\centering 
\begin{tabular}{r|rrrrrr}
& \multicolumn{6}{c}{$N=4000$, $N_\mathrm{pred}=500$} \\
  \hline
method & secs & {\footnotesize $\sqrt{1-\mathrm{NSE}}$} & RMSE & 95\%c & SD & $p$-val \\ 
  \hline
alc2 & 46.3 & 0.0196 & 0.884 & 1.000 & 2.71 & 0.0776 \\ 
  mspe2 & 83.8 & 0.0197 & 0.889 & 1.000 & 2.72 & 0.0000 \\ 
  nnbig & 20.0 & 0.0225 & 1.018 & 0.992 & 1.41 & 0.0000 \\ 
  alc & 23.5 & 0.0259 & 1.172 & 0.999 & 2.79 & 0.4164 \\ 
  mspe & 41.9 & 0.0259 & 1.172 & 1.000 & 2.78 & 0.0000 \\ 
  csc99 & \ \,3105.8 & 0.0309 & 1.396 & 0.961 & 1.47 & 0.0000 \\ 
  csc999 & 181.9 & 0.0337 & 1.525 & 0.957 & 1.54 & 0.0000 \\ 
  nn & 0.5 & 0.0647 & 2.920 & 0.964 & 2.60 & 0.0047 \\ 
  mspe.nomle & 41.3 & 0.0673 & 3.040 & 0.992 & 4.77 & 0.0022 \\ 
  alc.nomle & 23.2 & 0.0681 & 3.075 & 0.991 & 4.79 & 0.0000 \\ 
  nnbig.nomle & 8.8 & 0.0719 & 3.246 & 0.811 & 2.78 & 0.0000 \\ 
  nn.nomle & 0.2 & 0.1637 & 7.388 & 0.799 & 5.43 &  \\ 
   \hline
\end{tabular}
\caption{Average timings (in seconds) and accuracy (in $\sqrt{1-\mathrm{NSE}}$) values
  from  thirty-fold Monte Carlo experiments on the borehole data.
  The final column contains 
  $p$-values from one-sided $t$-tests of adjacent performers (better
  v.~next-best).  The rows of the Table are ordered by the accuracy
  estimate(s).}
\label{t:borehole}
\end{table}

Table \ref{t:borehole} summarizes the results of a comparison involving thirty
Monte Carlo repetitions, each with new random LHS testing and training sets.
Package defaults are used in all cases; second-stage greedy methods (e.g.,
``mspe2'') used unsmoothed first stage $\theta_x = \hat{\theta}_n(x)$ values
(from ``mspe''). The timings, obtained on the same $4$-core hyperthreaded iMac
throughout, comprise of the sum of fitting and prediction stages.  {\tt
OpenMP} pragmas, with 8 threads, were used to parallelize our local design
methods.  Although CSC is not explicitly multi-threaded, the sparse linear
algebra libraries it uses (from the {\tt spam} package) are lightly threaded
and were observed to utilize between one and five threads in this example.  We
can see clearly from the table(s) that even without parallelization ($\sim\!8$x
slower) our local methods are orders of magnitude faster than the CSC ones.

Differences in the raw accuracy numbers, via average $\sqrt{1 - \mathrm{NSE}}$
or RMSE, are much greater than they were in our earlier illustrative example.
Acknowledging the randomness in the Monte Carlo setup, the final column gives
a $p$-value for a one-sided paired $t$-test under the alternative that that
row's NSE came from a different population than those with the next-lowest
averages (next row down).  For example, at the 5\% level, the best method
(``alc2'') is not statistically better than the second best (``mspe2''),
whereas both are convincingly better than the third best, and so on.  At a
high level, observe that the local methods without $\hat{\theta}(x)$ fare
worst. NN $(n=50)$ is dominated by greedy methods across the board.  Due to
lack of statistical significance in differences between MSPE and ALC, the
former might not be worth the added computational expense.  As before, a big
NN ($n=200$) is competitive in terms of accuracy per second, however only when
local MLEs are calculated.  Without local inference and parallelization (both
our contributions),  the ``big'' NN method is amongst the most expensive
($8.8\times 8 \approx 70$ secs) and the least accurate.

Focusing on the uncertainty estimates in the table, observe that the greedy
methods over cover, whereas the CSCs have the correct nominal pointwise
coverage.  They also have, with the exception of ``nnbig'', lower
predicted standard deviation.  Interpreting these results is not
straightforward, however.  The true response is smooth and deterministic, so the
5\% which CSC mis-covers must comprise a small number of contiguous regions
of the input space where it was (possibly substantially) biased.  We therefore
draw comfort from the fact that the local methods are both more accurate and
more conservative.  The low standard deviation of ``nnbig'' would seem to be
due solely to the larger $n$ in the denominator of $\psi_n(x)$ in
(\ref{eq:gpk}), not to a better estimate of the global variance structure.
This can be verified by replacing $50\cdot\psi_n(x)/200$ in the coverage
calculations for the other $n=50$ estimators, which results in values nearly
identical to the others.  We conclude that our greedy methods are (perhaps
overly) conservative.  However next to ``nn'', which provides deceptively good
coverage results, caution in terms of predictive uncertainty pays dividends in
accuracy.


Results for a double-sized experiment are shown in Table \ref{t2:borehole}.
Accuracy results increase slightly, but relative orders are similar.  What is
noteworthy is that the local methods require, as expected, about double the
computational effort.  By contrast the CSC methods take four to eight times
longer, suggesting a quadratic scaling in $N$.  Anecdotally we remark that the
99\% sparse CSC method on $(N,n)=(8000, 1000)$ approaches the limit of the
size of problem possible on this machine due to memory constraints.  We tried
$N=10000$.  Frequent memory pages to disk lead to dramatic slowdowns; our
estimates suggest the code would have taken days to finish.  By contrast, the
local/greedy methods required seconds more.

\section{Discussion}
\label{sec:discuss}

We outlined a family of sequential design schemes to modernize an old idea of
approximate kriging based on local neighborhoods [for a recent reference, see
\citet{emory:2009}].  The result is a predictor that is fast, nonstationary,
highly parallelizable, and whose ``knobs'' offer direct control on speed
versus accuracy.  Really the only tuning parameter is $n$. We argue in favor
of calculating local designs which differ from a simpler (yet modernized) NN,
and give two greedy algorithms which have the same local $O(n^3)$
computational order as NN.  But the order notation hides large constant, which
means that the greedy methods demand compute cycles that could have been
spent, e.g., by NN with a larger $n$ budget. Indeed, our results show that
this approach is competitive, but not without modern enhancements like local
$\hat{\theta}_n(x)$ and parallel computation. Nevertheless, instinct suggests
that smaller $n$ is better. Many of our ideas for extension, which we outline
below, bear this out.

By saving a small amount of information---indices of the local designs at each
$x$, and corresponding $\hat{\theta}_n(x)$ values---designs can augmented,
picking up where they left off if more compute cycles become available.  Those
resources do not need to be allocated uniformly. Larger local designs $n(x)$
can be sought where, e.g., the potential to reduce variance or increase
accuracy is larger. Such decisions can be made more judiciously (i.e.,
accurately, via ALC) and efficiently (smaller $n$ calculations) under a greedy
local design scheme, compared to NN say. This is suggestive of benefit from
differential effort through the stages of global inference in Section
\ref{sec:param}.  One could start with a small $n$ and local inference based
on NN; then iterate with ALC and/or MSPE, possibly increasing $n$; and finally
allocate additional $n(x)$, refining hard-to-predict areas until the
computing budget is exhausted.

Having a stopping criterion might be desirable when computational limits are
less constrained, which when applied independently to obtain
$n_\mathrm{thresh}(x)$ for each $x$ could represent an alternative mechanism
for allocating computational resources differentially amongst the predictive
locations.  One option is to track the ratio of predictive and reduced
variances ($V_j(x)/V_{j+1}(x)$) over the iterations, and stop beyond a certain
minimum threshold.  Both depend on $\psi_j$, so cancellation would lead to a
metric independent of the responses $\{y_i\}_{i=1}^j$ (other than via a common
estimate of $\theta$). However, non-uniform full-data designs coupled with
spatially varying choices for $\theta$ make the actual {\em time} to
convergence---which is related cubically to $n_{\mathrm{thresh}}(x)$---highly
unpredictable.  Since lack of predictability would be compounded over
thousands of independent local searches, practical considerations may
necessitate a low global cap on $n_{\mathrm{thresh}}(x)$, recommending against
a simplistic NN approach.

We remark that there is a trade-off between the size of local designs and the
non-stationary flexibility of the global predictor: as the designs get larger,
the global surface can exhibit less spatial heterogeneity.  With larger $n$,
approximation will improve relative to the full-data GP, but it may exhibit
poorer out-of-sample performance, suggesting it could be beneficial to stop
early.  Post-process pruning steps, perhaps via an out-of-sample validation
scheme, could be deployed to find $n^*(x) \leq n$, a search which is more
efficient the smaller $n$ is to start with.  Another option is to augment the
GP with elaborate mean structure, following \citet{kaufman:etal:2012}. That
would allow a sparser covariance structure, which for us is a sparser local
design (again smaller $n$).  That mean structure is one of the big strengths
of the CSC method, which we anticipate would fare better than ours in
extrapolation exercises, or when predicting in an area of of the input space
with very sparse design coverage.

When the design and predictive grids are dense, and when the same $\theta$
parameters are used globally, the pattern of greedy local designs in the
interior of the input space (away from the boundaries) are highly regular. For
example, Figure \ref{f:alcmspe} suggests a simple template-based rule (e.g.,
some NNs and some farther out) may be a thrifty alternative to
expensive MSPE/ALC calculation. 
Along similar lines, we have found that substantial speedups can be obtained by searching
first over the closest members of $X-X_j(x)$.  One option is to stop
early once some tolerance is reached on changes to ALC/MSPE. Even simpler is
to restrict searches to a more modestly sized local neighborhood of candidates
$N' \ll N$, say $N' = n^m$ for $m$-dimensional input spaces.  Actually, for
all experiments in this paper, we restricted searches to the nearest
$N'=$1000 candidates. Results (except time) with $N'=10000$.

We envision further potential benefits from parallelization.
Numbers of desktop computing cores are doubling every few years and our methods are
well-positioned to take advantage. Graphical processing units (GPUs) are
taking off for scientific computing, having thousands of stripped-down
computing cores.  
An early early GPU version of
ALC--based local design search has yielded 20--70x efficiency gains
\citep{gramacy:niemi:weiss:2013}. 

Finally, we anticipate that a modern local approach to GP emulation will have
applicability beyond emulation.  For example, they may be ideal for computer
model calibration where predictions are only required over part of the input
space: at pairings of the calibration parameter(s) with a small number of
field data sites.  In that case it might be sensible to build ``local''
designs jointly for field data locations.  Although the details for doing this
by MSPE are still under development, an ALC version can be invoked by
providing a matrix of $x$-values to the {\tt laGP} function in the {\sf R}
package. Other contexts clearly pose challenges for local emulation. For
example, when optimizing black box functions by expected improvement---where
designs ($N$) are small and predictive grids are potentially enormous.

\subsection*{Acknowledgments}

The authors would like to thank Jarad Niemi for thoughtful discussions on
early drafts, and Brian Williams for a careful proofing of our mathematical
appendices. We would also like to thank two referees and an AE for
constructive criticism on our initial submission. This work was supported in
part by NSF grant number CMMI-1233403.

\pagebreak
\appendix
\renewcommand{\thesection}{SM \arabic{section}}
\renewcommand{\thesubsection}{SM\S\arabic{subsection}}
\setcounter{figure}{0}
\setcounter{table}{0}
\setcounter{equation}{0}
\renewcommand{\thefigure}{SM.\arabic{figure}}    
\renewcommand{\thetable}{SM.\arabic{table}}    
\renewcommand{\theequation}{SM.\arabic{equation}}    
\setcounter{page}{1}
\renewcommand{\thepage}{SM \arabic{page}}

\section*{Supplementary Materials}

\subsection{Partition inverse equations}
\label{sec:pie}

\citet{barnett:1979} gives the following decomposition.
\begin{align*}
\mbox{If} &&
K(x) &= \begin{bmatrix}
K & k(x) \\
k^\top(x) & K(x,x)
\end{bmatrix},
&& \mbox{then} &
K^{-1}(x) &= \begin{bmatrix}
[K^{-1} + g(x) g^\top(x) v(x)] & g(x) \\
g^\top(x) & v^{-1}(x)
\end{bmatrix},
\end{align*}
where $g(x) = -v^{-1}(x) K^{-1} k(x)$ and $v(x) = K(x,x) -
k^\top(x) K^{-1} k(x)$. 

\subsection{Mean-squared predictive error}
\label{sec:mspe}

Here we derive the expressions behind the MSPE (\ref{eq:mspe})
development in Section \ref{sec:loc}.  We allow any form of the
correlation function where derivatives are analytic.  Although our
expressions in the main body of text assume a scalar parameter
$\theta$, the derivations here allow for a $p$-parameter family, where
$p \geq 1$.  Therefore our partial derivatives are expressed
component-wise and vectorized as appropriate.  In most cases, the
single-parameter result is obtained straightforwardly by removing
subscripts and collapsing products of identical second derivatives
into squares and/or factors of two.

We begin with the batch log likelihood derivatives which lead to the
Fisher information.  Define $w_j = K^{-1}_j Y_j$, which is a component
of the marginal (log) likelihood (\ref{eq:gpk}) for data $D_j$ via
$\psi_j$.  Much of what follows repeatedly leverages that
\begin{equation}
\frac{\partial K_j^{-1}}{ \partial
\theta_k} = - K_j^{-1} \frac{\partial K_j}{\partial \theta_k} K_j^{-1},
\label{eq:dKi}
\end{equation}
where $\frac{\partial K_j}{\partial \theta_k}$ is the matrix of
partial derivatives corresponding to $K_j$ calculated via $K_{\theta,
  \eta}(x, x')$ for all $(x, x')$ pairs in $D_j$.  Recursive
application yields $\frac{\partial^2 K_j}{\partial \theta_k \partial
  \theta_l}$.  Then, for $1 \leq k,l \leq p$, we have
\begin{align}
\frac{\partial w_j}{\partial \theta_k} &= - K_j^{-1} \frac{\partial
  K_j}{\partial \theta_k} w_j, & 
\mbox{and } \;\;\; \frac{\partial^2 w_j}{\partial \theta_k \partial
  \theta_l} &= - K_j^{-1} \left[ \frac{\partial^2 K_j}{\partial
    \theta_k \partial \theta_l}w_j + \frac{\partial K_j}{\partial
    \theta_k} \frac{\partial w_j}{\partial \theta_l} + 
\frac{\partial K_j}{\partial
    \theta_l} \frac{\partial w_j}{\partial \theta_k} \right].
\label{eq:d2w}
\end{align}
Also useful is
\[
\frac{\partial \log | K_j|}{\partial \theta_k} =
\frac{1}{|K_j|} \frac{ \partial |K_j|}{\partial \theta_k} =
\frac{|K_j| \,\mathrm{tr}\!\left\{ K_j^{-1} \frac{\partial K_j}{\partial
      \theta_k}\right\}}{ | K_j| } =
\mathrm{tr}\!\left\{ K_j^{-1} \frac{ \partial K_j}{\partial \theta_k} \right\}.
\]
Note that the trace of a product of square matrices $M^{(1)}$ and
$M^{(2)}$ may be efficiently computed as $\sum_{k,l} M^{(1)}_{lk}
M^{(2)}_{kl}$.

Using the above results, the first and second derivatives of the log
likelihood $\ell_j(\theta) \equiv \log p(Y_j | \theta, X_j)$ given in
(\ref{eq:gpk}) are (for $1 \leq k,l \leq p$)
\begin{align*}
\frac{\partial \ell_j(\theta)}{\partial \theta_k} &= - \frac{1}{2} \,
\mathrm{tr}\!\left\{ K_j^{-1} \frac{\partial K_j}{\partial \theta_k}
\right\} + \frac{j Y_j^\top K_j^{-1} \frac{\partial
  K_j}{\partial \theta_k} K_j^{-1} Y_j }{2\psi_j},  \\
\mbox{and } \;\;\; \frac{\partial^2 \ell_j(\theta)}{\partial
  \theta_k \partial \theta_l} &= - \frac{1}{2} \,\mathrm{tr}\!\left\{
K_j^{-1} \left[ \frac{\partial^2 K_j}{\partial \theta_k \partial
    \theta_l} - \frac{\partial K_j}{\partial \theta_l}
  K_j^{-1} \frac{\partial K_j}{\partial \theta_k} \right] \right\} \\
&\;\;\; - \frac{j Y^\top K_j^{-1}}{2 \psi_j} \left[
2 \frac{\partial K_j}{\partial \theta_l}
  K_j^{-1} \frac{\partial K_j}{\partial \theta_k} -  \frac{\partial^2 K_j}{\partial \theta_k \partial
    \theta_l}  \right] K_j^{-1} Y_j \\
&\;\;\; + \frac{j}{2 \psi_j^2} \left(Y_j^\top K_j^{-1} \frac{\partial
  K_j}{\partial \theta_k} K_j^{-1} Y_j\right)\left(Y_j^\top K_j^{-1} \frac{\partial
  K_j}{\partial \theta_l} K_j^{-1} Y_j\right).
\end{align*}
The Fisher information $F_j(\theta)$ matrix is obtained by negating
the $(k,l)$ elements of the second derivative above.  Some simplifications
arise in the the one-dimensional
parameter case, which we quote below for easy reference when following
the development in the main manuscript.
\begin{align}
\ell'(\theta) &\equiv \frac{\partial \log p(Y | K)}{\partial \theta} = - \frac{1}{2}
\mathrm{tr}\left\{ K^{-1} \dot{K} \right \} + \frac{N}{2} \times \frac{
(Y^\top K^{-1} \dot{K} K^{-1} Y)}{\psi}, \label{eq:dl} \\
\ell''(\theta) &\equiv \frac{\partial^2 \log p(Y | K)}{\partial^2
  \theta} =
-\frac{1}{2} \mathrm{tr} \left\{ K^{-1} \left[ \ddot{K} - \dot{K}
    K^{-1} \dot{K}\right]\right\} \label{eq:d2l} \\
&+ \frac{N}{2 \psi} \times Y^\top K^{-1} \left[ \ddot{K} - 2 \dot{K} K^{-1}
  \dot{K}  \right] K^{-1} Y+ \frac{N}{2\psi^2} \times \left[ Y^\top K^{-1} \dot{K} K^{-1}
  Y\right]^2, \nonumber
\end{align}
where $\dot{K}$ and $\ddot{K}$ are shorthand for $\frac{\partial
  K}{\partial \theta}$ and $\frac{\partial^2 K}{\partial^2 \theta}$
respectively.  

Sequential updating of the Fisher information leverages a recursive
expression of the log likelihood which follows trivially from a
cascading conditional representation of the joint probability of the
responses given the parameters
$
\ell_{j+1}(\theta) = \log p(Y_{j+1} | \theta) = \log p(Y_j | \theta) +
\log p(y_{j+1} | Y_j, \theta) \equiv \ell_j(\theta) + \ell_j(y_{j+1};
\theta)
$ 
where the final term represents the conditional log-likelihood for
$y_{j+1}$ given $Y_j$ and $\theta$.  Taking (negative) second derivatives yields
the following updating equations
\[
\left\{F_{j+1}(\theta)\right\}_{kl} = - \frac{\partial^2 \ell_j(\theta)}{\partial
  \theta_k \partial \theta_l} - \frac{\partial \ell_{j}(y_{j+1};
  \theta)}{\partial \theta_k \partial \theta_l} = \{F_j(\theta)\}_{kl} - 
 \frac{\partial \ell_{j}(y_{j+1};
  \theta)}{\partial \theta_k \partial \theta_l}
\]
Deriving the final term, here, requires expressions for $y_{t+1}|Y_t,
\theta$, i.e., the predictive equations, and their derivatives.
First,
\begin{align*}
\ell_j(y_{t+1}; \theta) &= \mbox{const } - \frac{1}{2} \log
V_j(x_{j+1}; \theta) - \left( \frac{j+1}{2} \right) \log \left[ 
j-2+ \frac{(y_{j+1} - \mu_j(x_{j+1}; \theta))^2}{V_j(x_{j+1};
  \theta) }\right] \\
&\equiv \mbox{const } - \frac{1}{2} \log
V_j - \left( \frac{j+1}{2} \right) \log \left[ 
j-2+ \frac{(y_{j+1} - \mu_j)^2}{V_j}\right],  
\end{align*}
thereby defining some shorthand that will be useful going forward.

The derivatives of the log likelihood follow immediately from
applications of the chain rule.  For $1 \leq k, l \leq p$,
\begin{align}
& \frac{\partial \ell_j(y_{j+1}; \theta)}{\partial \theta_k} = -
\frac{1}{2V_j} \frac{\partial V_j}{\partial \theta_k} +
\left( \frac{j+1}{2} \right) \left[ j - 2 + \frac{(y_{j+1} -
    \mu_j)^2}{V_j} \right]^{-1} \label{eq:dlx} \\
&\;\;\;\; \times \left[ \frac{2(y_{j+1} - \mu_j)}{V_j}
  \frac{\partial \mu_j}{\partial \theta_k} + \frac{(y_{j+1} -
    \mu_j)^2}{V_j^2} \frac{\partial V_j}{\partial \theta_k}
\right]. \nonumber \\
&\frac{\partial^2 \ell_j(y_{j+1}; \theta)}{\partial \theta_k \partial
  \theta_l} =
- \frac{1}{2 V_j} \frac{\partial^2 V_j}{\partial \theta_k \partial
  \theta_l} + \frac{1}{2V_j^2} \frac{\partial V_j}{\partial \theta_k}
\frac{\partial V_j}{\partial \theta_l}  + \left( \frac{j+1}{2} \right) \left[ j-2 + \frac{(y_{j+1} -
     \mu_j)^2}{V_j} \right]^{-2}  \label{eq:d2lx} \\
&\;\;\;\; \times \left[ \frac{2(y_{j+1} -
     \mu_j)}{V_j} \frac{\partial \mu_j}{\partial\theta_k} + \frac{(y_{j+1} -
     \mu_j)^2}{V_j^2} \frac{\partial V_j}{\partial \theta_k} \right]
\left[ \frac{2(y_{j+1} -
     \mu_j)}{V_j} \frac{\partial \mu_j}{\partial \theta_l} + \frac{(y_{j+1} -
     \mu_j)^2}{V_j^2} \frac{\partial V_j}{\partial \theta_l} \right] \nonumber \\
&\;\;\;\; + \left( \frac{j+1}{2} \right) \left[ j-2 + \frac{(y_{j+1}
  - \mu_j)^2}{V_j} \right]^{-1} \left\{ - \frac{2}{V_j} \frac{\partial
\mu_j}{\partial \theta_k} \frac{ \partial \mu_j}{\partial \theta_l}
 - \frac{2(y_{j+1} - \mu_j)}{V_j^2} \left[ \frac{\partial \mu_j}{\partial
  \theta_k} \frac{\partial V_j }{ \partial \theta_l} + \frac{\partial \mu_j}{\partial
  \theta_l} \frac{\partial V_j }{ \partial \theta_k} \right] \right. \nonumber
\\
& \;\;\;\; \left. + \frac{2(y_{j+1} - \mu_j)}{V_j} \frac{\partial^2
  \mu_j}{\partial \theta_k \partial \theta_l} - \frac{2(y_{j+1} -
  \mu_j)^2}{V_j^3} \frac{ \partial V_j}{\partial \theta_k}
\frac{\partial V_j}{\partial \theta_l} + \frac{(y_{j+1} - \mu_j)^2}{V_j^2}
\frac{\partial^2 V_j}{\partial \theta_k \partial \theta_l} \right\} \nonumber
\end{align}

Eqs.~(\ref{eq:dlx}--\ref{eq:d2lx}) require the derivatives
of the predictive equations.  In turn, those require $w_j$ and its
derivatives (\ref{eq:d2w}), and a new quantity $z_j =
K_j^{-1} k_j \equiv K_j^{-1} k_j(x_{j+1}; \theta)$ and its derivatives:
\begin{align}
\frac{\partial z_j}{\partial \theta_k} &= K_j^{-1} \left[
  \frac{\partial k_j}{\partial \theta_k} - \frac{\partial
    K_j}{\partial \theta_k} z_j \right], \label{eq:dz} \\
 \mbox{and } \;\;\; \frac{\partial^2 z_j}{\partial \theta_k \partial \theta_l} &= K_j^{-1}
\left[ 
\frac{\partial^2 k_j}{\partial \theta_k \partial \theta_l} -
  \frac{\partial^2 K_j}{\partial \theta_k \partial \theta_l} z_j -
  \frac{\partial K_j}{\partial \theta_k} \frac{\partial z_j}{\partial
    \theta_l} - \frac{\partial K_j}{\partial \theta_l} \frac{\partial
    z_j}{\partial \theta_k} \label{eq:d2z}
\right]
\end{align}
which follow from applications of (\ref{eq:dKi}).
The derivatives of the mean are then revealed as
\begin{align}
\frac{\partial \mu_j}{\partial \theta_k} &= \frac{\partial
  z_j^\top}{\partial \theta_k} Y_j, & &\mbox{and}&
\frac{\partial^2 \mu_j}{\partial \theta_k \partial \theta_l} &= 
\frac{\partial^2 z_j^\top}{\partial \theta_k \partial \theta_l} Y_j. \label{eq:dmu}
\end{align}

For the variances it helps to write $V_j = \frac{1}{j-2} \times \psi_j
\times v_j$ where $v_j \equiv v_j(x_{j+1}; \theta) = K_\eta(x_{j+1},
x_{j+1}) - k_j^\top(x_{j+1}) K_j^{-1} k_j(x_{j+1})$, and $\psi_j
\equiv \psi_j(\theta)$.  We assume that when the arguments to
$K_{\theta, \eta}$ are identical, as in $v_j$, the correlation
function is no longer a function of $\theta$.  Derivatives then follow
as
\begin{align}
\frac{\partial V_j}{\partial \theta_k} &= \left(\frac{1}{j-2}\right) \left[
\frac{\partial \psi_j}{ \partial \theta_k}  v_j + \psi_j \frac{\partial
  v_j}{\partial \theta_k}\right] \nonumber \\
&= \left(\frac{1}{j-2}\right) \left[ Y_j^\top \frac{\partial
  w_j}{\partial \theta_k} v_j - \psi_j \left(\frac{\partial
    k_j^\top}{\partial \theta_k} z_j + k_j^\top \frac{\partial
    z_j}{\partial \theta_k} \right) \right], \;\;\; \mbox{ and} \label{eq:dV} \\
\frac{\partial ^2 V_j}{\partial \theta_k \partial
  \theta_l} &=
\left( \frac{1}{j-2} \right) \left\{ 
Y_j^\top \frac{\partial^2 w_j}{\partial \theta_k \partial \theta_l}
v_j + Y_j^\top \frac{\partial w_j}{\partial \theta_k} \frac{\partial
  v_j}{\partial \theta_l} - \frac{\partial \psi_j}{\partial \theta_l}
\left[ \frac{\partial k_j^\top}{\partial \theta_k} z_j + k_j^\top
  \frac{\partial z_j}{\partial \theta_k} \right] \right.  \nonumber \\
& \;\; - \left. \psi_j \left[ \frac{\partial^2
      k_j^\top}{\partial \theta_k \partial \theta_l} z_j + \frac{\partial
      k_j^\top}{\partial \theta_k} \frac{\partial z_j}{ \partial
      \theta_l} + \frac{\partial k_j^\top}{\partial
      \theta_l} \frac{\partial z_j}{\partial \theta_k} + k_j^\top
    \frac{\partial^2 z_j}{\partial \theta_k \partial \theta_l} \right]
\right\} \nonumber \\
&= \left( \frac{1}{j-2} \right) 
\left\{ 
Y_j^\top \frac{\partial^2 w_j}{\partial \theta_k \partial \theta_l}
 v_j - Y_j^\top \frac{\partial w_j}{\partial \theta_k}
 \left[\frac{\partial k_j^\top}{\partial \theta_l} z_j + k_j^\top
 \frac{\partial z_j}{\partial \theta_l} 
 \right] \right. \label{eq:d2V} \\
&\;\; - Y_j^\top \frac{\partial w_j}{\partial \theta_l}
  \left[ \frac{\partial k_j^\top}{\partial \theta_k} z_j + k_j^\top
    \frac{\partial z_j}{\partial \theta_k} \right] \left. - \, \psi_j
    \! \left[ 
\frac{\partial^2 k_j^\top}{\partial \theta_k \partial \theta_l} z_j +
\frac{\partial k_j^\top}{\partial \theta_k} \frac{\partial
  z_j}{\partial \theta_l} + \frac{\partial k_j^\top}{\partial \theta_l} \frac{\partial
  z_j}{\partial \theta_k}  + k_j^\top \frac{\partial^2 z_j}{\partial
  \theta_k \partial \theta_l} \right] \right\}. \nonumber
\end{align}
This completes the expressions required for updating the Fisher
information matrix.

The MSPE calculation (\ref{eq:mspe}) requires that the conditional
likelihood be replaced with the expectation
\[
\mathcal{G}_{j+1}(\theta) = F_j(\theta) + \bE\left\{-\frac{\partial^2
    \ell_j(y_{j+1}; \theta)}{\partial\theta_k \partial\theta_l}  \right \}
\]
because we wish to select the next $x_{j+1}$ in a manner that does not
utilize the ``future'' observation $y_{j+1}$.  Unfortunately, the
Student-$t$ predictive equations (\ref{eq:predgp}--\ref{eq:preds2})
preclude a tractable analytic expectation calculation.  Therefore, we
approximate by employing Gaussian surrogate equations
with matched moments.  I.e.,
\begin{equation}
\ell_j(y_{j+1}; \theta) = \log p(y_{j+1} | Y_j, \theta) \approx 
-\frac{1}{2} \log{2\pi} - \frac{1}{2} \log V_j - \frac{(y_{j+1} -
  \mu_j)^2}{2V_j}.
\label{eq:lapprox}
\end{equation}
The (exact) derivatives for the approximation (\ref{eq:lapprox}) then
follow, for $1 \leq k,l, \leq p$
\begin{align}
\frac{\partial \ell_j(y_{j+1}; \theta)}{\partial \theta_k} &\approx -
\frac{1}{2V_j} \frac{ \partial V_j}{\partial \theta_k} +
\frac{(y_{j+1} - \mu_j)}{V_j} \frac{\partial \mu_j}{\partial \theta_k}
+ \frac{(y_{j+1} - \mu_j)^2}{2V_j^2} \frac{\partial V_j}{\partial
  \theta_k}, \label{eq:dla} \\
\mbox{and } \;\; \frac{\partial^2 \ell_j(y_{j+1}; \theta)}{\partial
  \theta_k \partial \theta_l} &\approx - \frac{1}{2V_j}
\frac{\partial^2 V_j}{\partial \theta_k \partial \theta_l} +
\frac{1}{2V_j^2} \frac{\partial V_j}{\partial \theta_k} \frac{\partial
  V_j}{\partial \theta_l} - \frac{1}{V_j} \frac{\partial
  \mu_j}{\partial \theta_k} \frac{\partial \mu_j}{\partial \theta_l}
\label{eq:d2la} \\
&\;\;\; + \frac{(y_{j+1} - \mu_j)}{V_j}
\frac{ \partial^2\mu_j}{\partial \theta_k \partial \theta_l}
- \frac{(y_{j+1} - \mu_j)}{V_j^2} \left [ \frac{\partial \mu_j}{\partial
  \theta_k} \frac{\partial V_j}{ \partial \theta_l} 
+ \frac{\partial V_j}{\partial
  \theta_k} \frac{\partial \mu_j}{ \partial \theta_l}  \right]
\nonumber \\
& \;\;\; + \frac{(y_{j+1}
  - \mu_j)^2}{ 2 V_j^2} \frac{ \partial^2 V_j}{\partial
  \theta_k \partial \theta_l} - \frac{(y_{j+1} - \mu_j)^2}{V_j^3}
\frac{\partial V_j}{\partial \theta_k} \frac{\partial V_j}{\partial \theta_l}.
\nonumber 
\end{align}
Taking the (negative) expectation of (\ref{eq:d2la}) gives,
\begin{align}
\bE \left \{-\frac{\partial^2 \ell_j(y_{j+1}; \theta)}{\partial
  \theta_k \partial \theta_l} \Big{|} Y_j, \theta \right \} &=
\frac{1}{2V_j} \frac{\partial^2 V_j}{\partial \theta_k \partial
  \theta_l} - \frac{1}{2V_j^2} \frac{\partial V_j}{\partial
  \theta_k} \frac{\partial V_j}{\partial \theta_l} + \frac{1}{V_j}
\frac{\partial \mu_j}{\partial \theta_k} \frac{\partial
  \mu_j}{\partial \theta_l} \nonumber \\
&\;\;\; - \frac{1}{2V_j} \frac{\partial^2 V_j}{\partial
  \theta_k \partial \theta_l} + \frac{1}{V_j^2} \frac{\partial
  V_j}{\partial \theta_k} \frac{\partial V_j}{\partial \theta_l}
\nonumber \\
&= \frac{1}{2V_j^2} \frac{\partial
  V_j}{\partial \theta_k} \frac{\partial V_j}{\partial \theta_l} 
+ \frac{1}{V_j}
\frac{\partial \mu_j}{\partial \theta_k} \frac{\partial
  \mu_j}{\partial \theta_l} \label{eq:Ed2la}
\end{align}

Now, given a specified input site $x$ at which predictions of $Y(x)$
are desired, and given data $D_j$ containing previously selected
sites, the objective is to choose the next site $x_{j+1}$ recognizing
that the data pairs will lead to improved predictions via
(\ref{eq:predgp}--\ref{eq:preds2}) {\em and} an improved estimate of
the correlation parameters $\theta$.  If the latter were of primary
concern, one possible way to choose $x_{j+1}$ would be to maximize the
determinant of $\mathcal{G}_{j+1}(\theta)$ for some choice of $\theta$.  While
reasonable, this does not directly consider the prediction error
variance of $Y(x)$.  Therefore it possess no mechanism for giving a
local character to the estimate of $\theta$.  So instead we propose to
choose $x_{j+1}$ to minimize the predictive variance of $Y(x)$ in a
manner that considers the impact of estimation uncertainty in
$\theta$.  Specifically, we choose $x_{j+1}$ to minimize the Bayesian
mean squared prediction error defined as
\begin{equation}
J(x_{j+1}, x) = \bE \left\{ \left[Y(x) - \mu_{j+1}(x;
  \hat{\theta}_{j+1})\right]^2\Big{|} Y_j \right\}, \label{eq:mspe1}
\end{equation}
where $\mu_{j+1}(x; \hat{\theta}_{j+1})$ depends on $(x_{j+1},
y_{j+1})$ and represents the hypothetical prediction of $Y(x)$ that
would result after choosing $x_{j+1}$, observing $y_{j+1}$,
calculating MLE $\hat{\theta}_{j+1}$ based on the expanded data set
$D_{j+1}$, and using the MLE in the prediction of $Y(x)$.
Eq.~(\ref{eq:mspe1}) becomes
\begin{align}
J(x_{j+1}, x) &= \bE \left \{ \left[ Y(x) - \mu_{j+1}(x; \theta) +
    \mu_{j+1}(x; \theta) - \mu_{j+1}(x; \hat{\theta}_{j+1}) \right]^2
  \Big{|} Y_j \right\} \nonumber \\
&= \bE \left \{ \bE \left \{ \left[ Y(x) - \mu_{j+1}(x; \theta) + \mu_{j+1}(x; \theta) - \mu_{j+1}(x;
\hat{\theta}_{j+1}) \right]^2 \Big{|} Y_j, y_{j+1},
    \theta \right\} \Big{|} Y_j \right\} \nonumber \\
&= \bE \left \{ \bE \left \{ \left[ Y(x) - \mu_{j+1}(x; \theta)  \right]^2 \Big{|} Y_j, y_{j+1},
    \theta \right\} \Big{|} Y_j \right\} \nonumber \\
&\;\;\; + 2 \bE \left \{ \bE \left \{ \left[ Y(x) - \mu_{j+1}(x; \theta)
    \right]\left[ \mu_{j+1}(x; \theta) - \mu_{j+1}(x;
\hat{\theta}_{j+1}) \right] \Big{|} Y_j, y_{j+1},
    \theta \right\} \Big{|} Y_j \right\} \nonumber \\
&\;\;\; +  \bE \left \{ \bE \left \{ \left[   \mu_{j+1}(x; \theta) - \mu_{j+1}(x;
\hat{\theta}_{j+1}) \right]^2 \Big{|} Y_j, y_{j+1},
    \theta \right\} \Big{|} Y_j \right\} \nonumber \\
&= \bE \left\{ V_{j+1}(x; \theta) \Big{|} Y_j \right\} + 0 + \bE
\left\{ \left[ \mu_{j+1}(x; \theta) - \mu_{j+1}(x; \hat{\theta}_{j+1})
  \right]^2 \Big{|} Y_j \right \}. \label{eq:mspe2}
\end{align}

Regarding the first term in (\ref{eq:mspe2}), we write it as
\begin{align}
\bE \left\{ V_{j+1}(x; \theta) \Big{|} Y_j \right\}  = 
\bE \left\{ \frac{\psi_{j}(\theta)}{(j-1)} v_{j+1}(x ; \theta) \Big{|} Y_j\right\}
& = \frac{\psi_j(\theta)}{j-2} v_{j+1}(x; \theta) \label{eq:mspe2a} \\
& \approx \frac{\psi_j}{j-2} v_{j+1}(x; \hat{\theta}_j) \nonumber
\end{align}
which is the new/reduced variance (\ref{eq:dxy}) after adding $x_{j+1}$ into
$D_j$, marginalized with respect to the unknown $y_{j+1}$.  The last
approximate equality in (\ref{eq:mspe2a}) results from substituting the
unknown $\theta$ by its estimate calculated form $D_j$.  The second equality
in (\ref{eq:mspe2a}) follows readily from Eq.~(\ref{eq:psiup}), which can be
written as (omitting $\theta$ for brevity)
\begin{align*}
\psi_{j+1} &= \psi_j + \frac{\mu_j^2(x_{j+1})}{v_j(x_{j+1})} 
- \frac{2 y_{j+1} \mu_j(x_{j+1})}{v_j(x_{j+1})}
+ \frac{y_{j+1}^2}{v_j(x_{j+1})} 
= \psi_j + \frac{[y_{j+1} - \mu_j(x_{j+1})]^2}{v_j(x_{j+1})}, \\
\intertext{so that}
\bE\{\psi_{j+1} | Y_j\} &= \psi_j + \frac{V_j(x_{j+1})}{v_j(x_{j+1})} 
= \psi_j + \frac{\psi_j v_j(x_{j+1})}{(j-2) v_j(x_{j+1})} = \frac{(j-1) \psi_j}{(j-2)}.
\end{align*}
Regarding the last term in (\ref{eq:mspe2}), we again approximate as
follows.
\begin{align}
\mu_{j+1}(x; \theta) - \mu_{j+1}(x; \hat{\theta}_{j+1}) &\approx \left[
  \frac{\partial \mu_{j+1}(x; \theta)}{\partial \theta}
  \Big{|}_{\theta = \hat{\theta}_{j+1}} \right]^\top (\theta -
  \hat{\theta}_{j+1}) \nonumber \\
& \approx
 \left[
  \frac{\partial \mu_{j}(x; \theta)}{\partial \theta}
  \Big{|}_{\theta = \hat{\theta}_{j}} \right]^\top (\theta -
  \hat{\theta}_{j+1}) \label{eq:mspe2b}
\end{align}
The last inequality in (\ref{eq:mspe2b}) involves two approximations.
First, the partial derivative of $\mu_{j+1}(x; \theta)$ at future
$\hat{\theta}_{j+1}$ is approximated by the partial derivative at the
current $\hat{\theta}_j$.  The second approximation uses the partial
derivative of $\mu_j(x; \theta)$ instead.  Both are required for
tractability because $\hat{\theta}_{j+1}$ and $\mu_{j+1}(x, \theta)$
depend on the future $y_{j+1}$ in a manner that would make the
expectation in (\ref{eq:mspe2a}) difficult to evaluate analytically.

Note that the difference between $\mu_{j+1}(x; \theta)$ and $\mu_j(x,
\theta)$, and their partial derivatives at a common value of $\theta$,
will generally be similar for larger $j$ because the most influential
points are included in initial iterations.  For situations in which
$y_{t+1}$ is available in advance, one might consider using it for
calculating the partial derivative of $\mu_{j+1}(x; \theta)$, and
thereby avoid an approximated expectation.  However, even though this
would lend tractability and accuracy in one sense, it may not be as
attractive a criterion as using $\mu_j(x , \theta)$.  A strategy that
used the partial derivative of $\mu_{j+1}(x; \theta)$ might avoid
adding a site at which $y_{j+1}$ was not consistent with the other
$\{y_i, i = 1, 2, \dots, j\}$, which would ignore unusual or
unpredictable features of the response surface.

Substituting (\ref{eq:mspe2a}) and (\ref{eq:mspe2b}) into
(\ref{eq:mspe2}), and assuming the posterior covariance matrix of
$\theta$ about its MLE can be approximated by the inverse of the
Fisher information matrix $\mathcal{G}_{j+1}(\hat{\theta}_j)$ given in
(\ref{eq:mspe}), leaves us with
\begin{align*}
J(x_{j+1}, x) &= \frac{\psi_j}{j-2} v_{j+1}(x; \hat{\theta}_j) \\
&\;\;\; + \bE
\left\{ \left[ \frac{\partial \mu_{j}(x; \theta)}{\partial \theta}
  \Big{|}_{\theta = \hat{\theta}_{j}} \right]^\top (\theta - \hat{\theta}_{j+1})(\theta -
  \hat{\theta}_{j+1})^\top
\left[ \frac{\partial \mu_{j}(x; \theta)}{\partial \theta}
  \Big{|}_{\theta = \hat{\theta}_{j}} \right] \Big{|} Y_j \right\}\\
&\approx \frac{\psi_j}{j-2} v_{j+1}(x; \hat{\theta}_j)  +
\left[ \frac{\partial \mu_{j}(x; \theta)}{\partial \theta}
  \Big{|}_{\theta = \hat{\theta}_{j}}
\right]^\top \mathcal{G}_{j+1}^{-1}(\hat{\theta}_j) 
\left[ \frac{\partial \mu_{j}(x; \theta)}{\partial \theta}
  \Big{|}_{\theta = \hat{\theta}_{j}} \right].
\end{align*}

\subsection{Comparison of condition numbers for local designs}
\label{sec:condum}

A knock-on effect of the spacing of the sub-design points found by the
greedy method is that the condition numbers (ratio of the largest to
smallest eigenvalue)\footnote{We used the {\tt kappa} function in {\sf
    R} with default arguments.} of the resulting covariance matrices,
$K_j$, are lower relative to those of the NN method as $j$ becomes
large.
\begin{figure}[ht!]
\centering
\includegraphics[scale=0.5, trim=10 5 5 10]{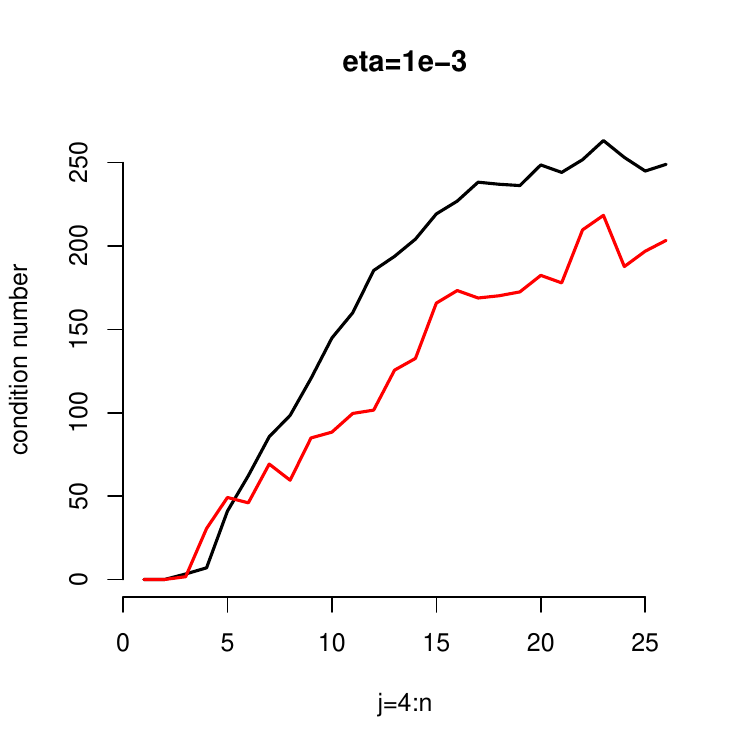}
\hspace{0.5cm}
\includegraphics[scale=0.5, trim=10 5 5 10]{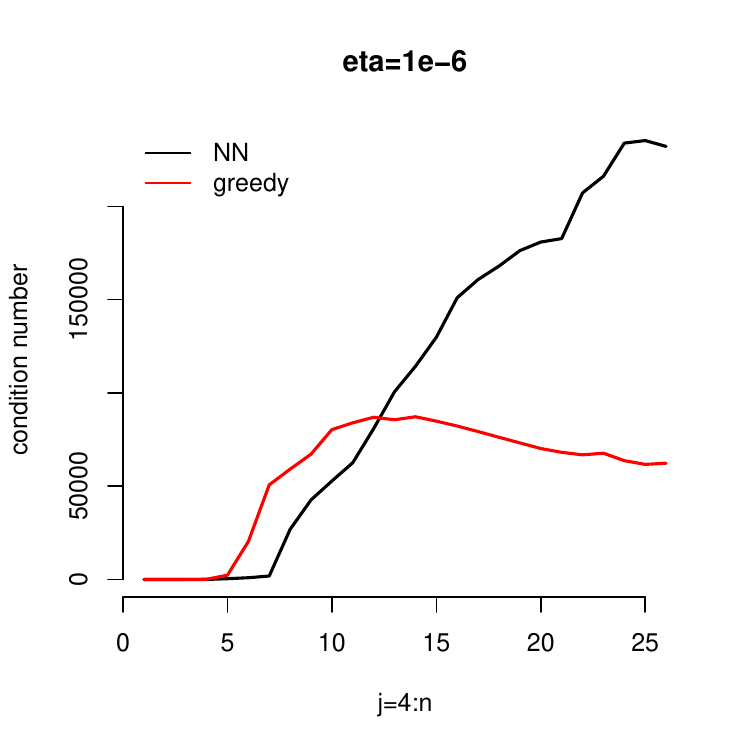}
\caption{Condition numbers, {\tt kappa}$(K_j^{-1})/\mbox{\tt dim}(K_j^{-1})^2$,
  versus design size, $j$, 
  for NN and greedy methods for two different values of the
  nugget, $\eta$.}
\label{f:condn}
\end{figure}
Figure \ref{f:condn} provides an illustration.  The {\em left} plot
corresponds to exactly the setup giving rise to the {\em right} panel in
Figure \ref{f:alcmspe}, using ALC; whereas the {\em right} plot uses a
smaller nugget value, $\eta$.  Lower condition numbers indicate
greater numerical stability and thus more reliable matrix
decompositions.  This is relevant for GP surrogate modeling since it
means that smaller nuggets, and thus truer interpolations, can be
achieved with the greedy method if so desired.  Although some recent
papers have argued that the quest for truer interpolation in surrogate
modeling may not be efficient from a statistical perspective
\citep{gra:lee:2012,peng:wu:2012}, seeking out the smallest possible
nugget while retaining numerical stability remains an aspiration in
the literature \citep{ranjan:haynes:karsten:2011}.

\subsection{An interpretation of the reduction in variance criterion in
  terms of partial correlations}
\label{sec:pcor}

The ALC criterion (\ref{eq:dxy}) has a helpful interpretation in terms
of the partial correlation between the prediction errors at $x$ and at
some $x'$, a potential new site to be added into the design.  For the
case when all parameters are known, denote the error in predicting
$Y(x)$, given data $D_j$, by $e(x | Y_j) = Y(x) - \mu_j(x; \theta)$.
Note that $e(x | Y_j)$ and $Y_j$ are orthogonal under the standard
correlation inner product, and consider the orthogonal decomposition
\begin{equation}
Y(x) = \mu_j(x; \theta) + e(x|Y_j) = \mu_j(x; \theta) + \rho
\frac{\sqrt{\Var[e(x | Y_j)]}}{\sqrt{\Var [e(x' | Y_j)]}} e(x'|Y_j) +
e(x| Y_j, x') \label{eq:od}
\end{equation}
where 
\[
\rho = \Corr[e(x | Y_j),  e(x' | Y_j)] = \frac{\Cov[e(x|Y_j),  e(x' |
  Y_j)]}{\sqrt{\Var[e(x | Y_j)]}\sqrt{\Var[e(x' | Y_j)]}}
\]
denotes the correlation coefficient between the errors in predicting
$Y(x)$ and $Y(x')$.  Sometimes this is called the {\em partial
  correlation coefficient} between $Y(x)$ and $Y(x')$ given $Y_j$.

Because of the orthogonality of the three terms in (\ref{eq:od}), the
error variance in predicting $Y(x)$ given $Y_j$ is
\begin{equation}
\Var[e(x|Y_j)] = \rho^2 \Var[e(x|Y_j)] + \Var[e(x | Y_j, x')]. \label{eq:ev}
\end{equation}
The right-most term in (\ref{eq:ev}) is defined as the error variance
in predicting $Y(x)$ after including the response at $x'$.  Hence, the
term
\begin{equation}
\rho^2 \Var[ e(x | Y_j)] = \Corr^2[ e(x | Y_j), e(x' | Y_j)] \Var[e(x |
Y_j)] = \frac{\Cov^2 [ e(x | Y_j), e(x' | Y_j)]}{\Var[e(x' | Y_j)]} \label{eq:dxy2}
\end{equation}
is the reduction in predictive variance in $Y(x)$ after including the
response at $x'$.

Noting that $
\Cov[e(x | Y_j), e(x' | Y_j)] = \tau^2[ K(x, x') - k_j^\top(x) K_j^{-1} k_j(x')]$,
it is straightforward to show that (\ref{eq:dxy2}) is equivalent to the
criterion in (\ref{eq:dxy}) with $\psi_j/(j-2)$ estimating $\tau^2$.  Because
$\Var[e(x | Y_j)]$ does not depend on $x'$, the expression (\ref{eq:dxy2}) for
the reduction in variance implies that the next site $x_{j+1}$ should be
chosen to maximize the correlation between the errors in predicting $Y(x)$ and
$Y(x_{j+1})$ given $Y_j$.

\subsection{Tabular results for empirical comparisons}
\label{sec:tabs}

\begin{table}[ht!]
\centering
\begin{tabular}{r|rrrrrr}
& \multicolumn{6}{c}{$N=8000$, $N_\mathrm{pred}=1000$} \\
  \hline
method & secs & {\footnotesize $\sqrt{1-\mathrm{NSE}}$} & RMSE & 95\%c & SD & $p$-val \\ 
  \hline
mspe2 & 169.1 & 0.0173 & 0.787 & 1.000 & 2.49 & 0.0490 \\ 
  alc2 & 92.3 & 0.0174 & 0.790 & 1.000 & 2.49 & 0.0000 \\ 
  nnbig & 37.2 & 0.0201 & 0.911 & 0.993 & 1.25 & 0.0000 \\ 
  mspe & 84.8 & 0.0230 & 1.046 & 1.000 & 2.56 & 0.0056 \\ 
  alc & 46.2 & 0.0233 & 1.058 & 1.000 & 2.57 & 0.0000 \\ 
  csc99 & 26519.5 & 0.0279 & 1.267 & 0.965 & 1.40 & 0.0000 \\ 
  csc999 & 2814.1 & 0.0328 & 1.490 & 0.959 & 1.53 & 0.0000 \\ 
  nnbig.nomle & 18.3 & 0.0545 & 2.472 & 0.789 & 2.07 & 0.1618 \\ 
  mspe.nomle & 84.1 & 0.0548 & 2.486 & 0.989 & 3.60 & 0.0000 \\ 
  alc.nomle & 45.1 & 0.0558 & 2.533 & 0.985 & 3.61 & 0.0000 \\ 
  nn & 1.3 & 0.0622 & 2.822 & 0.954 & 2.32 & 0.0000 \\ 
  nn.nomle & 0.6 & 0.1425 & 6.470 & 0.750 & 4.29 &  \\ 
\end{tabular}
\caption{Extending Table \ref{t:borehole} with a double-sized experiment; 
average timings (in seconds) and accuracy (in $\sqrt{1-\mathrm{NSE}}$) values
  from thirty-fold Monte Carlo experiments on the borehole data. 
  The final column  contains
  $p$-values from one-sided $t$-tests of adjacent performers (better
  v.~next-best).  The rows of the table are ordered by the accuracy
  estimate(s).}
\label{t2:borehole}
\end{table}

Table \ref{t2:borehole} contains results on the borehole data with
an experiment of twice the size as the one reported in the main document
in Table \ref{t:borehole}.  The accuracy comparison is very similar,
but note that the CSC comparators require almost $10\times$ time whereas
our greedy local approximations runtimes scale linearly.
Both tables report a variant of Nash--Sutcliffe efficiency
\citep[NSE,][]{nash:sutcliffe:1970} to connect with results reported
in the CSC paper.
\begin{equation}
\mathrm{NSE} = 1 - \frac{\sum_{x \in X_\mathrm{pred}} (\hat{Y}(x) -
  Y(x))^2}{\sum_{x \in X_\mathrm{pred}} (Y(x) - \bar{Y})^2} \label{eq:nse}
\end{equation}
Here, $\hat{Y}(x)$ represents the predicted value of $Y(x)$, which for all
models under comparison is the posterior predictive mean; $\bar{Y}$ is
the average value of $Y(x)$ over the design space.  The second term in
(\ref{eq:nse}) is estimated mean-squared predictive error (i.e., realized
$\mathrm{RMSE}^2$, that which we approximately optimize over when calculating
a local MSPE design) over the unstandardized variance of $Y(x)$. NSE is an
attractive metric because it can be interpreted as analog of $R^2$. However,
it has the disadvantage of providing misleadingly large values in
deterministic (noise-free) computer model emulation contexts.  We therefore
report $\sqrt{1-\mathrm{NSE}} = \mathrm{sd}(\mbox{prediction
error})/\mathrm{sd}(\mbox{response})$ instead, and in addition to other
metrics like RMSE.  NSEs can still be backed out for direct comparison to
tables in the CSC paper.

\begin{table}[ht]
\centering
\vspace{0.25cm}
\begin{tabular}{rrrrrrr}
  \hline
 method & secs & {\footnotesize $\sqrt{1-\mathrm{NSE}}$} & RMSE & 95\%c & SD \\
  \hline
mspe2 & 1434.8 & 0.0041 & 0.0008 & 1.0 & 0.0058 \\ 
  alc2 & 829.0 & 0.0041 & 0.0008 & 1.0 & 0.0058 \\ 
  nnbig & 603.7 & 0.0050 & 0.0010 & 1.0 & 0.0027 \\ 
  mspe & 717.5 & 0.0050 & 0.0010 & 1.0 & 0.0060 \\ 
  alc & 406.2 & 0.0050 & 0.0010 & 1.0 & 0.0060 \\ 
  nnbig.nomle & 198.6 & 0.0055 & 0.0011 & 1.0 & 0.0036 \\ 
  mspe.nomle & 683.6 & 0.0064 & 0.0013 & 1.0 & 0.0063 \\ 
  alc.nomle & 401.4 & 0.0064 & 0.0013 & 1.0 & 0.0063 \\ 
  nn & 32.3 & 0.0114 & 0.0023 & 1.0 & 0.0045 \\ 
  nn.nomle & 26.3 & 0.0120 & 0.0024 & 1.0 & 0.0048 \\ 
   \hline
   \hline
\end{tabular}
\caption{Average timings (in seconds), accuracy (in $\sqrt{1-\mathrm{NSE}}$
  and RMSE, smaller is better), and uncertainty (95\% coverage and predicted
   standard deviation) on a single out-of-sample  experiment.  The
  timings for two-stage estimators (``mspe2'' and ``alc2'') include those from
  the first stage (``mspe'' and ``alc'').  A ``nomle'' indication flags
  estimates based on $\theta_0 = 1$.}
\label{t:compare}
\end{table}

Table \ref{t:compare} summarizes results of a similar comparison on our
illustrative 2d example.  The experiment is described in more detail in
Section \ref{sec:rill}.

\end{document}